\documentclass[twocolumn,twocolappendix,numberedappendix,appendixfloats]{openjournal}

\usepackage[utf8]{inputenc}
\usepackage[T1]{fontenc}
\usepackage{float}
\usepackage{xcolor}
\usepackage{graphicx}
\usepackage{amsmath}
\usepackage{amssymb}
\usepackage{booktabs}
\usepackage{enumitem}
\usepackage{placeins}
\usepackage{hyperref}

\hypersetup{colorlinks=true,
	urlcolor=blue,
	linkcolor=blue,
	citecolor=blue,
    menucolor=blue}

\def\apjs{{\rm ApJS}}

\def\kms{\, {\rm km}\, {\rm s}^{-1}}
\def\lya{Ly$\alpha$}
\newcommand{\GammaHI}{\Gamma_{\mathrm{HI}}}

\newcommand{\astrid}{{\tt Astrid}}

\newcommand{\ZThreeCatalogueBias}{5.59}
\newcommand{\ZThreeCatalogueBiasError}{^{+0.42}_{-0.45}}
\newcommand{\ZThreeChiNoFloor}{10.33}
\newcommand{\ZThreeClusteringBias}{5.593}
\newcommand{\ZThreeClusteringChi}{0.65}
\newcommand{\ZThreeClusteringGamma}{1.376}
\newcommand{\ZThreeClusteringMinus}{0.382}
\newcommand{\ZThreeClusteringPlus}{0.528}
\newcommand{\ZThreeClusteringRaw}{0.964}
\newcommand{\ZThreeDESIFloorHat}{0.0120}
\newcommand{\ZThreeEffectiveBias}{3.883}
\newcommand{\ZThreeEffectiveChi}{0.85}
\newcommand{\ZThreeEffectiveFloorHat}{0.0135}
\newcommand{\ZThreeEffectiveGamma}{1.422}
\newcommand{\ZThreeEffectiveRaw}{1.471}
\newcommand{\ZThreeMMedian}{-25.894}
\newcommand{\ZThreeNASTRID}{762}
\newcommand{\ZThreeNDESI}{6{,}245}
\newcommand{\ZThreeNTemplates}{48}
\newcommand{\ZThreeReqMedian}{1.77}
\newcommand{\ZThreeZEffective}{2.976}
\newcommand{\ZThreeZMedian}{2.975}

\newcommand{\ZThreeClusteringSamplingMinus}{0.076}
\newcommand{\ZThreeClusteringSamplingPlus}{0.095}
\newcommand{\ZThreeClusteringSystematicMinus}{0.521}
\newcommand{\ZThreeClusteringSystematicPlus}{0.837}
\newcommand{\ZThreeClusteringExpandedTotalMinus}{0.524}
\newcommand{\ZThreeClusteringExpandedTotalPlus}{0.847}
\newcommand{\ZThreeEffectiveSamplingMinus}{0.078}
\newcommand{\ZThreeEffectiveSamplingPlus}{0.098}
\newcommand{\ZThreeEffectiveSystematicMinus}{0.537}
\newcommand{\ZThreeEffectiveSystematicPlus}{0.863}

\newcommand{\zfigure}[2][\textwidth]{\includegraphics[width=#1]{updated_figures/#2.pdf}}

\begin{document}

\title{Measuring the ultraviolet background at $z\simeq3$ from the quasar proximity effect in DESI DR1}

\author{Patrick Shaw$^{1,*}$}
\author{Rupert A.C. Croft$^{1}$}
\thanks{$^*$E-mail: pshaw2@andrew.cmu.edu}
\affiliation{$^{1}$ McWilliams Center for Cosmology, Dept. of Physics,
Carnegie Mellon University, Pittsburgh, PA 15213, USA
\\
\\
}

\begin{abstract}
Ionizing radiation from quasars reduces the abundance of neutral hydrogen in the surrounding intergalactic medium (IGM), suppressing Ly$\alpha$ absorption. This absorption deficit, known as the quasar proximity effect, can be used to measure the ultraviolet background (UVBG). We measure the hydrogen photoionization rate of the UVBG in Data Release 1 of the Dark Energy Spectroscopic Instrument (DESI). Our sample contains \ZThreeNDESI{} quasars over $2.75\leq z<3.25$, selected with rest-frame absolute magnitudes at 1450\,\AA\ of $M_{1450}\leq-25.5$ and signal-to-noise ratios of at least 3 measured redward of Ly$\alpha$. We reconstruct quasar continua with principal component analysis trained on Hubble Space Telescope spectra and stack the resulting transmitted-flux profiles. We fit the resulting stacked profile with a halo model that accounts for reduced absorption from quasar radiation and enhanced absorption in the surrounding overdense environment. We constrain the environmental contribution using independent measurements of quasar clustering bias, while the transmission profile constrains the UVBG. To validate our method, we construct mock spectra matched to the luminosity and noise properties of our DESI sample using the \astrid{} hydrodynamical simulation at $z=3$. We apply the same analysis pipeline to the mock spectra and DESI data, testing the recovery of known input photoionization rates and using the results to calibrate the final UVBG measurement.
We obtain
\[
\GammaHI=\left(\ZThreeClusteringGamma^{+\ZThreeClusteringSamplingPlus}_{-\ZThreeClusteringSamplingMinus}\right)\times10^{-12}\,\mathrm{s}^{-1}
\]
at $z\simeq3$. The statistical errors are obtained by bootstrap resampling the DESI sightlines. We also estimate a systematic uncertainty of $^{+\ZThreeClusteringSystematicPlus}_{-\ZThreeClusteringSystematicMinus}\times10^{-12}\,\mathrm{s}^{-1}$.

\end{abstract}


\maketitle

\section{Introduction}
\label{sec:intro}

Radiation from a quasar raises the hydrogen photoionization rate in the surrounding intergalactic medium (IGM) above the level maintained by the ultraviolet background (UVBG). The resulting suppression of Lyman-alpha (Ly$\alpha$) absorption near the quasar is known as the proximity effect. This effect was first recognized by \citet{Carswell1982} and placed on a quantitative footing by \citet{Bajtlik1988}. The formalism established there underpins most subsequent measurements \citep{Lu1991,Bechtold1994,Scott2000,DallAglio2008LOS,DallAglio2008}. At fixed gas density and temperature, photoionization equilibrium implies that the neutral-hydrogen abundance, and hence the Ly$\alpha$ optical depth, varies inversely with the total hydrogen photoionization rate \citep{GunnPeterson1965}. The strength of the proximity effect therefore depends on the ratio of the quasar's ionizing flux to that of the UVBG. Because the quasar contribution falls approximately as the inverse square of distance, that ratio is encoded in the radial extent of the enhanced transmission. Given an estimate of the quasar's luminosity, the proximity profile can therefore constrain the background hydrogen photoionization rate, $\Gamma_{\rm HI}$.

Accurate measurements of $\Gamma_{\rm HI}$ are scientifically important for understanding the ionization state of the IGM and the sources that maintain it. Near $z\simeq3$, quasars and star-forming galaxies both contribute ionizing radiation, but their relative contributions remain uncertain, in part because the escape fraction of ionizing photons from galaxies is poorly constrained \citep{HaardtMadau2012,KhaireSrianand2019,Puchwein2019,FaucherGiguere2020}. Combined with measurements of the ionizing-photon mean free path and an assumed spectral shape, $\Gamma_{\rm HI}$ constrains the total ionizing emissivity of quasars and galaxies \citep{BeckerBolton2013}. Because $\Gamma_{\rm HI}$ also controls the neutral-hydrogen abundance, it affects the mean transmitted forest flux and therefore the cosmological inferences drawn from Ly$\alpha$ forest observations \citep{Kim2007,Adame2025}. Most existing constraints over $2<z<5$ infer the background from the mean forest opacity and an assumed thermal history \citep{BeckerBolton2013}. The proximity effect provides a complementary constraint through the transmission enhancement around quasars, with a different dependence on continuum reconstruction, quasar radiation, and the surrounding gas.

For most of its history the proximity effect was measured with tens to hundreds of sightlines \citep{Bechtold1994,Cooke1997,Scott2000,DallAglio2008}. The Sloan Digital Sky Survey (SDSS) expanded the available quasar samples substantially \citep{Dawson2013}. Using 1,733 quasars from SDSS Data Release 5, \citet{DallAglio2009} measured the proximity effect over $2.3<z<4.6$. The Dark Energy Spectroscopic Instrument (DESI) has expanded the available sample further, with approximately 1.6 million spectroscopically classified quasars in its first data release \citep{DESIDR1}. At this scale, precision is no longer limited by sample size alone. Coherent percent-level errors in continuum estimation, calibration, and modeling can survive spectral stacking and become increasingly important.

One complication in modeling the proximity effect is the overdense environment of quasar hosts. Quasars reside in massive, biased dark-matter halos and therefore occupy regions denser than the cosmic mean \citep{FontRibera2013}. The additional absorption associated with this overdensity partially offsets the transmission enhancement produced by the quasar's radiation. If it is neither modeled nor otherwise mitigated, this environmental contribution can be misinterpreted as a stronger UVBG and bias the inferred $\Gamma_{\rm HI}$ high \citep{LoebEisenstein1995,Rollinde2005,Guimaraes2007,Jalan2021}. Conversely, with assumptions about the UVBG and quasar radiation, absorption near quasars can be used to constrain their host-halo masses and surrounding overdensities \citep{FaucherGiguere2008,KimCroft2008,Hada2025}. In this work, we account for the quasar host environment with the halo model of \citet{Croft2025}, which represents the overdensity and quasar radiation simultaneously through the quasar bias parameter $b_q$ and the radiation equality radius $r_{\rm eq}$. For the DESI measurement, we use independent clustering measurements \citep{Charles2026} to constrain $b_q(z)$ and infer $\Gamma_{\rm HI}$ from the resulting proximity-effect transmission profile.

In this paper, we present our method for measuring the UVBG from the quasar proximity effect and apply it to a single redshift interval, $2.75\leq z<3.25$. We focus on $z\simeq3$ because this is the redshift at which we have validated the model and calibrated the recovery of the photoionization rate using matched \astrid{} mock spectra. Measurements at other redshifts, together with the corresponding validation and calibration, will be presented in a forthcoming paper.

The paper is organized as follows. Section~\ref{sec:model} summarizes the proximity-effect halo model and the conversion between the radial profile and $\Gamma_{\rm HI}$. Section~\ref{sec:data} describes DESI Data Release 1 (DR1) and our quasar sample selection. Section~\ref{sec:continuum} describes continuum reconstruction. Section~\ref{sec:method} describes how we fit the stacked transmission profiles to estimate $\Gamma_{\rm HI}$ and apply external-forest mean-flux regulation to correct the reconstructed continua. Section~\ref{sec:astrid} presents the validation and calibration using matched mock spectra. Results, uncertainty propagation, and interpretation follow in Sections~\ref{sec:results}--\ref{sec:summary_discussion}. Distances and luminosities are computed in a flat cosmology with a cosmological constant and cold dark matter ($\Lambda$CDM), with $H_0=67.7\,\mathrm{km\,s^{-1}\,Mpc^{-1}}$ and $\Omega_{\rm m}=0.31$.

\section{Modeling the proximity effect}
\label{sec:model}

\subsection{Transmitted flux and quasar bias}

The proximity-effect observable we measure is the transmitted flux,
\begin{equation}
F(\lambda)=\frac{f_{\rm obs}(\lambda)}{C(\lambda)},
\end{equation}
where $f_{\rm obs}(\lambda)$ is the measured quasar flux density and $C(\lambda)$ is the estimated unabsorbed quasar continuum. Radiation from the quasar reduces the neutral-hydrogen abundance in the surrounding gas, producing enhanced transmission near the quasar. We model the mean transmitted-flux profile, $F(r)$, where the line-of-sight separation $r$ is inferred from wavelength using the quasar redshift and adopted cosmology.

We use the cold dark matter (CDM) proximity-effect halo model developed by \citet{Croft2025}. It is built in four stages. First, the mean matter-density profile around a quasar host is modeled as the combination of a one-halo contribution and a large-scale two-halo term controlled by the quasar bias $b_q$. Second, a distribution of densities is assigned at each radius and mapped to a distribution of \lya\ optical depths. Third, the local ionizing radiation from the quasar is added to the background radiation field. Its relative strength is parameterized by the radiation equality radius $r_{\rm eq}$, the distance at which the quasar and background contribute equally to the hydrogen photoionization rate. Finally, peculiar velocities and thermal broadening are included to predict the transmission profile in redshift space.

\begin{figure}[t]
    \centering
    \zfigure[\columnwidth]{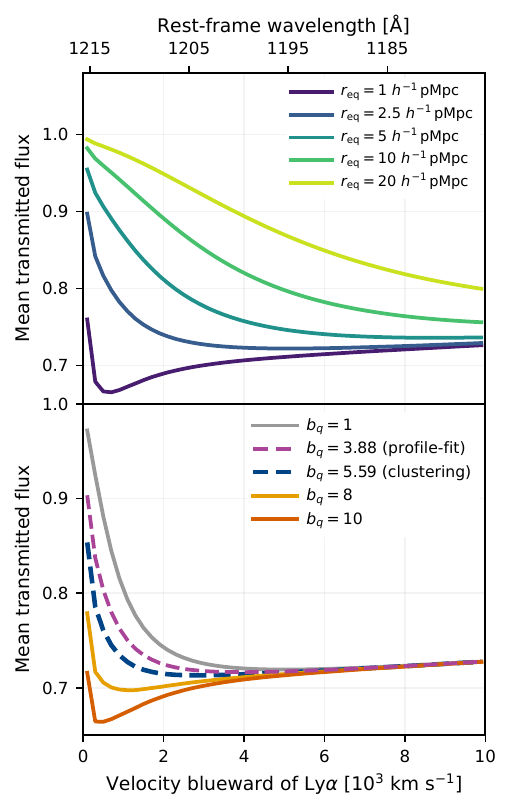}
    \caption{Examples of the halo-model proximity-effect profiles at $z=3$. The upper panel varies the radiation equality radius $r_{\rm eq}$ at fixed $b_q=5.6$, approximately the bias adopted for our sample. The lower panel varies $b_q$ at fixed $r_{\rm eq}=\ZThreeReqMedian\,h^{-1}\,\mathrm{pMpc}$, a representative equality radius for the primary sample. The curves include the mean-opacity evolution described in Section~\ref{sec:model_extensions}. The upper and lower horizontal scales show rest-frame wavelength and line-of-sight velocity separation, respectively. At $z=3$, $10{,}000\,\mathrm{km}\,\mathrm{s}^{-1}$ corresponds to approximately $22.1\,h^{-1}\,\mathrm{pMpc}$ of Hubble-flow proper separation in the adopted cosmology.}
    \label{fig:theory_profile_family}
\end{figure}

At a given redshift, each model transmission profile is specified by two principal parameters: the quasar bias $b_q$, which controls the overdensity of the surrounding environment, and the radiation equality radius $r_{\rm eq}$, which encodes the quasar's ionizing luminosity relative to the ultraviolet background. These parameters have opposing effects: increasing $b_q$ strengthens environmental absorption and lowers the transmitted flux, whereas increasing $r_{\rm eq}$ extends the region in which quasar radiation suppresses absorption and raises the transmitted flux. Figure~\ref{fig:theory_profile_family} illustrates these competing effects. We summarize the halo model below; a full description is given by \citet{Croft2025}.

\subsection{From density to transmitted flux}

Following the standard halo-model decomposition \citep{NeymanScott1952,Seljak2000,CooraySheth2002}, we adopt the quasar--matter cross-correlation of \citet{Croft2025},
\begin{equation}
\xi_{q\rho}(r)
=
\frac{A_{\rm NFW}\delta_c}{cr(1+cr)^2}
+
b_q\xi_{\rm CDM}(r),
\end{equation}
where
\begin{equation}
\delta_c
=
\frac{200}{3}
\frac{c^3}{\ln(1+c)-c/(1+c)}.
\end{equation}
The first term has a Navarro--Frenk--White (NFW) form \citep{Navarro1997} and describes the internal halo structure. We retain the notation and radial convention of \citet{Croft2025}, with concentration parameter $c$ and amplitude $A_{\rm NFW}$.

The two-halo term describes large-scale clustering, where $\xi_{\rm CDM}(r)$ is the linear-theory CDM correlation function computed using the Code for Anisotropies in the Microwave Background (CAMB; \citealt{LewisChallinor2011}), and $b_q$ is the linear quasar bias. The corresponding mean density in units of the cosmic mean is
\begin{equation}
\bar{\rho}(r)=1+\xi_{q\rho}(r).
\end{equation}

We map density to optical depth using the fluctuating Gunn--Peterson approximation \citep{Croft1997FGPA,Weinberg1997},
\begin{equation}
\tau(r)\propto\rho(r)^{\,2-0.7\gamma}\Gamma(r)^{-1},
\label{eq:fgpa}
\end{equation}
where $\rho(r)$ is the density in units of the cosmic mean, $\gamma$ is the exponent of the effective temperature--density relation, $T\propto\rho^\gamma$, and $\Gamma(r)$ is the total local hydrogen photoionization rate. The latter includes both the background rate, $\Gamma_{\rm HI}$, and the contribution from the quasar, and approaches $\Gamma_{\rm HI}$ far from the quasar. Following \citet{Croft2025}, we adopt $\gamma=0$, corresponding to an isothermal temperature--density relation. The temperature--density relation and observational motivation for this approximation near $z\simeq3$ are discussed by \citet{HuiGnedin1997} and \citet{Bolton2008}.

Because both $\tau$ and $F=\exp(-\tau)$ are nonlinear in density, the mean transmission profile cannot be obtained from $\bar{\rho}$ alone. Following \citet{Croft2025}, we adopt a lognormal density distribution $P(\rho\mid r)$ at each radius \citep{BiDavidsen1997,ColesJones1991}. Its mean follows the halo profile, $\bar{\rho}(r)=1+\xi_{q\rho}(r)$, and its density dispersion is a fixed fraction of the mean, $\sigma_\rho(r)=f\bar{\rho}(r)$. We adopt the value $f=1.25$ inferred from the \astrid{} simulation by \citet{Croft2025}. The effective optical depth is then
\begin{equation}
\tau_{\rm eff}(r)
=
-\ln\left[
\int_0^\infty P(\rho\mid r)
\exp[-\tau(\rho,r)]\,d\rho
\right].
\label{eq:taueff}
\end{equation}

At this stage, we retain the mean-transmission normalization of \citet{Kim2007} used in the halo-model calculations of \citet{Croft2025}. The subsequent mean-opacity evolution and external-forest continuum correction use the relation of \citet{Turner2024}, as described in Sections~\ref{sec:model_extensions} and \ref{sec:meanflux}, respectively.

\subsection{Quasar radiation and the equality radius}
\label{sec:quasar_radiation}

On the radial scales fitted here, which are well inside the expected ionizing-photon mean free path at $z\simeq3$ \citep{TheunsChan2024}, the quasar increases the local photoionization rate approximately as
\begin{equation}
\frac{\Gamma(r)}{\GammaHI}
=
1+\left(\frac{r_{\rm eq}}{r}\right)^2,
\label{eq:gammatot}
\end{equation}
where $r_{\rm eq}$ is the proper radius at which the quasar and UVBG contribute equally to the hydrogen photoionization rate. To calculate this radius from the measured $M_{1450}$, we adopt the luminosity-to-photon-rate conversion given by \citet{Croft2025},
\begin{equation}
\log_{10}\!\left[\frac{\dot N_\gamma}{\mathrm{s}^{-1}}\right]
=-0.4M_{1450}+46.366,
\end{equation}
where $\dot N_\gamma$ is the total emission rate of hydrogen-ionizing photons from the local quasar alone, excluding the UVBG contribution and assuming an escape fraction of unity.

We adopt an ionizing spectrum $L_\nu\propto\nu^{-\alpha}$ with $\alpha=1.5$. Because the hydrogen photoionization cross-section decreases with photon frequency, the total photon rate must be weighted by the cross-section when calculating its ionizing effect. Using $\sigma_\nu=\sigma_0(\nu/\nu_0)^{-3}$, where $\nu_0$ is the hydrogen ionization threshold and $\sigma_0=6.3\times10^{-18}\,\mathrm{cm}^{2}$, the photon-number-weighted mean cross-section is $\sigma_0\alpha/(\alpha+3)$. The radiation equality radius is therefore
\begin{equation}
r_{\rm eq}
=
\left[
\frac{\dot N_\gamma\sigma_0}{4\pi\GammaHI}
\frac{\alpha}{\alpha+3}
\right]^{1/2}.
\label{eq:req}
\end{equation}
We express the resulting proper radius in $h^{-1}\,\mathrm{pMpc}$. This expression includes the spectral weighting of the photoionization cross-section explicitly.

\subsection{Redshift-space distortions}
\label{sec:redshift_space}

The halo-model profiles are initially constructed in real space. Following the treatment of \citet{Croft1999}, coherent infall toward the quasar environment is modeled using the spherical-infall model of \citet{Yahil1985}, while small-scale peculiar velocities are described by an exponential pairwise-velocity distribution. Thermal broadening is included using a Gaussian approximation to the line profile. Mapping the real-space optical-depth distributions into redshift space with these velocity contributions and averaging the resulting transmission yields the redshift-space transmitted-flux profile.

\subsection{Mean-opacity evolution}
\label{sec:model_extensions}

The preceding calculations follow the halo model developed and tested by \citet{Croft2025}. Because that work used a simulation snapshot at a single redshift, the mean opacity of the model profiles does not evolve with redshift along the line of sight. For our observed proximity profiles, we must account for the evolution of the mean forest opacity across the absorption redshifts they span. We incorporate this evolution by rescaling the model optical depth according to the cosmic mean effective optical depth,
\begin{equation}
\bar{\tau}_{\rm eff}(z)
=
2.46\times10^{-3}(1+z)^{3.62},
\label{eq:mean_opacity_evolution}
\end{equation}
following \citet{Turner2024}. If $F_0(r)$ is the profile evaluated at the quasar redshift $z_q$, the evolved profile is
\begin{equation}
F(r)
=
\exp\left\{
-
\left[-\ln F_0(r)\right]
\frac{\bar{\tau}_{\rm eff}[z_{\rm abs}(r)]}
     {\bar{\tau}_{\rm eff}(z_q)}
\right\},
\label{eq:profile_redshift_evolution}
\end{equation}
where $z_{\rm abs}(r)$ is the absorption redshift corresponding to each location along the profile. This correction accounts for the smooth evolution of the mean forest opacity along the sightline while retaining the underlying halo model. The profiles in Figure~\ref{fig:theory_profile_family} include this opacity evolution.

\section{DESI DR1 sample}
\label{sec:data}

\subsection{DESI survey}

The Dark Energy Spectroscopic Instrument (DESI) provides a large, widely distributed sample of quasar spectra for measuring intergalactic absorption. DESI is mounted on the Mayall 4-m telescope at Kitt Peak National Observatory \citep{DESI2016}. Its first data release (DR1) includes the first 13 months of main-survey observations, covering more than $9{,}000\,\mathrm{deg}^2$ and yielding reliable redshifts for approximately 1.6 million spectroscopically classified quasars \citep{DESIDR1}. The spectral coverage of approximately $3600$--$9800$\,\AA\ includes both the \lya\ forest and the red-side regions used to reconstruct the continua of quasars near $z=3$. The large parent sample allows us to apply our sample-selection criteria within a narrow redshift interval while retaining thousands of sightlines for a stacked proximity-effect measurement. Its broad sky coverage also provides a basis for future comparisons between spatially separated quasar samples to investigate large-scale variations in the UVBG.

\subsection{Proximity-effect sample selection}
\label{sec:sample_selection}

\begin{table}
\centering
\caption{Selection criteria and properties of our DESI DR1 proximity-effect sample. The final sample also requires at least 30 usable pixels in the external forest at $1041$--$1160$\,\AA\ and a successful mean-flux correction as defined in Section~\ref{sec:meanflux}. The red-side signal-to-noise statistic is defined in Equation~\ref{eq:snr}.}
\label{tab:sample}
\small
\begin{tabular}{@{}ll@{}}
\toprule
Property & Selection or value\\
\midrule
\multicolumn{2}{@{}l}{\textit{Selection criteria}}\\
Redshift & $2.75\leq z<3.25$\\
Absolute magnitude & $M_{1450}\leq-25.5$\\
Red-side signal-to-noise & $(S/N)_{\rm red}\geq3$\\
Red-side continuum fit & Reduced $\chi^2\leq5$\\
Known broad-absorption-line quasars & Excluded\\
\midrule
\multicolumn{2}{@{}l}{\textit{Final sample}}\\
Number of quasars & \ZThreeNDESI\\
Median redshift & \ZThreeZMedian\\
Median absolute magnitude & $\ZThreeMMedian$\\
\bottomrule
\end{tabular}
\end{table}

We restrict our measurement to $2.75\leq z<3.25$, centered on the $z=3$ \astrid{} snapshot used to construct the matched mocks for our UVBG calibration (Section~\ref{sec:astrid}). Within this interval, we select luminous quasars with sufficient spectral quality to reconstruct their continua and measure the proximity effect. At fixed $\Gamma_{\rm HI}$ and quasar spectral shape, the quasar contribution to the local photoionization rate scales linearly with luminosity, while the characteristic proximity scale follows $r_{\rm eq}\propto L_{1450}^{1/2}$ (Equation~\ref{eq:req}). More luminous quasars therefore produce stronger and more extended proximity effects, making the transmission enhancement easier to measure. High red-side signal-to-noise provides better constraints on the continuum reconstruction.

We begin with quasars that have valid spectral products and usable Legacy Survey photometry, excluding known broad-absorption-line quasars. We adopt the updated quasar redshifts from the DESI DR1 Ly$\alpha$ forest redshift catalog, determined using modified Redrock quasar templates \citep{Bault2025}, when available; otherwise, we use the primary DESI redshift. For each object, we estimate $M_{1450}$ from extinction-corrected $g$, $r$, and $z$ photometry, using the band whose effective wavelength lies closest to $1450(1+z)$\,\AA. We calculate a power-law $K$-correction \citep{Hogg2002}, assuming $f_\nu\propto\nu^{-0.5}$ \citep{VandenBerk2001}, and require $M_{1450}\leq-25.5$.

Our spectral-quality selection requires a measured red-side signal-to-noise ratio $(S/N)_{\rm red}\geq3$ and a reduced $\chi^2\leq5$ for the red-side continuum fit. The continuum-fitting procedure and its validation are described in Section~\ref{sec:continuum}. Both quality statistics are determined entirely from spectral regions redward of \lya, so absorption fluctuations in the proximity region do not directly influence these cuts.

We measure $(S/N)_{\rm red}$ from the flux and inverse variance in three fixed rest-frame wavelength intervals: $1270$--$1292$, $1320$--$1355$, and $1445$--$1480$\,\AA. These intervals avoid the strongest emission features, including Ly$\alpha$/N\,V, the O\,I/Si\,II blend near $1305$\,\AA, Si\,IV/O\,IV] near $1400$\,\AA, and C\,IV near $1549$\,\AA. We retain pixels with finite flux and finite, positive inverse variance, requiring at least ten usable pixels across the combined windows. The statistic is the median per-pixel signal-to-noise ratio:
\begin{equation}
(S/N)_{\rm red}
=
\mathrm{median}\left[
f_{\rm obs}(\lambda)\sqrt{\mathrm{ivar}_\lambda}
\right].
\label{eq:snr}
\end{equation}
Here $\mathrm{ivar}_\lambda$ is the inverse variance of the measured flux density.

\begin{figure*}[!t]
\centering
\zfigure[\textwidth]{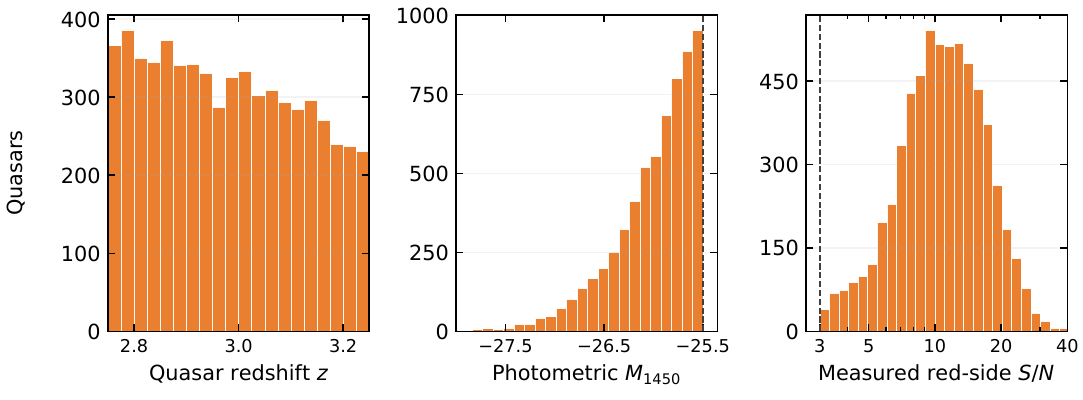}
\caption{Properties of the final DESI DR1 proximity-effect sample. The panels show the distributions of quasar redshift (left), photometric $M_{1450}$ (middle), and measured red-side signal-to-noise ratio (right). The sample spans $2.75\leq z<3.25$. Dashed lines mark the luminosity and signal-to-noise selection thresholds.}
\label{fig:sample}
\end{figure*}

After these cuts and the external-forest requirements described in Section~\ref{sec:meanflux}, the final sample contains \ZThreeNDESI{} quasars. Table~\ref{tab:sample} summarizes the sample, and Figure~\ref{fig:sample} shows its redshift, luminosity, and red-side signal-to-noise distributions. We combine the spectra into a mean transmission profile using weights proportional to $(S/N)_{\rm red}^2$, with the signal-to-noise capped at 10 when calculating the weights to prevent a small number of spectra from dominating. The construction and fitting of this profile are described in Section~\ref{sec:method}.

\section{Continuum fitting}
\label{sec:continuum}

The transmitted flux is obtained by dividing the observed spectrum by an estimate of the unabsorbed quasar continuum. Previous proximity-effect analyses have estimated this continuum by fitting splines directly through the absorption region blueward of \lya\ \citep{DallAglio2009,Jalan2021}. Fitting the continuum directly through the proximity region allows the fit to absorb the transmission enhancement we seek to measure. Corrections derived from mock spectra \citep{DallAglio2009} depend on how accurately the mocks reproduce the absorption and its effect on continuum placement. Predicting the continuum from the red-side spectrum avoids this source of signal suppression and is therefore better suited to measuring the proximity effect.

Methods that predict the blue-side continuum using only red-side spectral information include principal component analysis (PCA) and neural networks \citep{Suzuki2005,Lee2012,LiuBordoloi2021,Turner2024}. These methods use correlations between the red- and blue-side spectral shapes learned from training spectra. One such neural network is the Lyman-$\alpha$ Continuum Analysis Network (LyCAN), developed by \citet{Turner2024} for continuum reconstruction in the \lya\ forest. LyCAN was optimized using prediction errors over $1040$--$1200$\,\AA, whereas our measurement is particularly sensitive to systematic errors closer to the \lya\ emission peak. Accuracy in the more distant forest does not establish the accuracy needed in the proximity region. We therefore develop a PCA reconstruction tailored to the proximity region rather than directly adopting existing continuum estimates. Using the low-redshift Hubble Space Telescope/Cosmic Origins Spectrograph (HST/COS) reference continua assembled by \citet{Turner2024}, we construct the training set, adjust the predicted emission-line profile using red-side information, and evaluate the residual errors across the proximity region. We compare these predictions with LyCAN and use matched mock spectra to test and correct the complete continuum-fitting procedure.

\begin{figure*}[t!]
    \centering
    \zfigure[\textwidth]{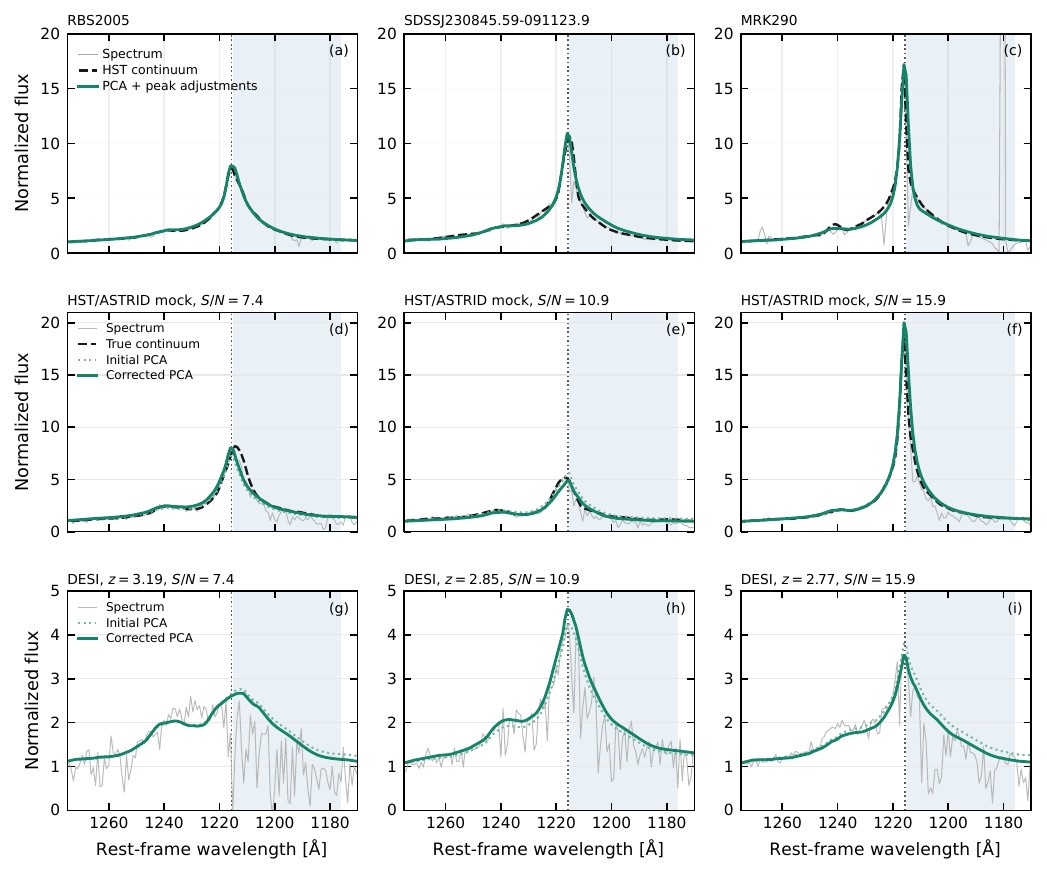}
    \caption{
    Representative continuum reconstructions using our PCA method. The first row shows HST/COS spectra and leave-one-out PCA predictions with the \lya\ emission-line adjustments described in Section~\ref{sec:continuum_desi}. The second row shows mock spectra constructed from HST continua and \astrid{} absorption. The third row shows DESI spectra. The mock and DESI examples are selected near the 20th, 50th, and 80th percentiles of our sample's red-side signal-to-noise distribution. Thin gray curves show the observed or noisy spectra, and black dashed curves show reference or true continua where available. Solid green curves show the adjusted PCA predictions in the HST row and the final corrected continua in the mock and DESI rows, including the corrections described in Sections~\ref{sec:meanflux}--\ref{sec:final_normalization}. Pale-green dotted curves show the initial PCA predictions in the mock and DESI rows. Pale-blue shading marks $200$--$9900\,\mathrm{km\,s^{-1}}$ blueward of \lya, where we fit the proximity effect. The vertical dotted line marks the \lya\ rest wavelength.
    }
    \label{fig:continuum_examples}
\end{figure*}

\begin{figure}[t!]
    \centering
    \zfigure[\columnwidth]{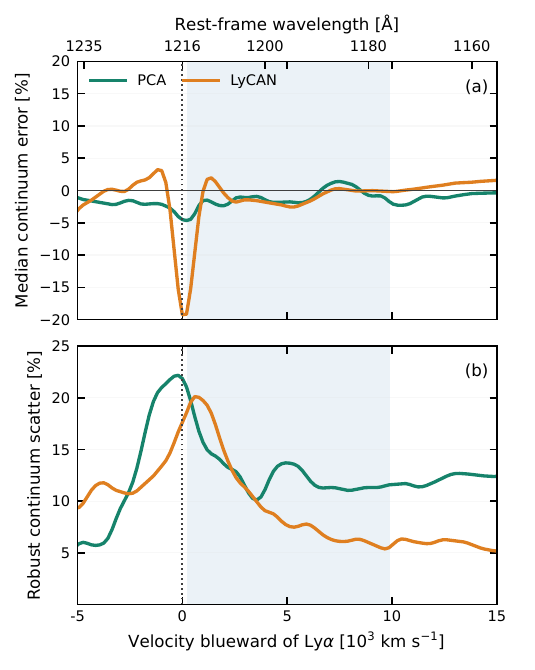}
    \caption{
    Continuum-prediction errors for PCA (green) and LyCAN (orange) on the 48 HST/COS objects. The upper panel shows the median fractional continuum error at each wavelength. The lower panel shows the object-to-object scatter, defined as half the difference between the 84th and 16th percentiles. The PCA predictions exclude each test object and its derived training continua from the training set and omit the \lya\ emission-line adjustments, as described in Section~\ref{sec:continuum_tests}. The supplied LyCAN estimates do not use an equivalent exclusion procedure. Velocity increases toward shorter wavelengths and the observer; rest-frame wavelength is shown along the upper axis. Pale-blue shading marks the $200$--$9900\,\mathrm{km\,s^{-1}}$ region used to fit the proximity effect.
    }
    \label{fig:continuum_method_comparison}
\end{figure}

\subsection{HST/COS training continua}
\label{sec:continuum_training}

We construct the PCA model using continuum fits to 48 low-redshift quasar spectra observed with the Cosmic Origins Spectrograph on the Hubble Space Telescope (HST/COS). These spectra were drawn from the Hubble Spectroscopic Legacy Archive \citep{Peeples2017HSLA} and independently continuum-fitted by \citet{Turner2024}. At these low redshifts, the sparse \lya\ forest leaves much of the intrinsic quasar spectrum directly visible, allowing the continuum on both sides of \lya\ to be fitted with substantially less ambiguity than in our DESI spectra. We use these existing fits as training references from which the PCA learns to predict the blue-side continuum using red-side information. Where the HST spectra lack blue-end coverage, the reference continua include the extrapolation described by \citet{Turner2024}.

Our DESI quasars are more luminous than the HST training quasars and are therefore expected to have weaker emission-line equivalent widths because of the Baldwin effect \citep{Baldwin1977,Dietrich2002}. To include this variation in the training set, we create additional versions of the HST continuum fits with weaker emission lines while preserving the underlying spectral shape. Each continuum is first placed on a common rest-frame grid of 699 pixels spanning $1040.8$--$1599.2$\,\AA\ and normalized by its median flux over $1264$--$1296$\,\AA. We estimate the continuum underlying the emission lines by fitting a cubic polynomial to logarithmic continuum flux as a function of $\ln\lambda$ in six relatively line-free intervals.\footnote{The intervals are $1085$--$1105$, $1140$--$1180$, $1270$--$1292$, $1320$--$1355$, $1445$--$1480$, and $1560$--$1585$\,\AA.} Exponentiating the polynomial gives this underlying continuum, which we subtract from the HST continuum fit to isolate the emission-line component.

For each HST continuum fit, we multiply the emission-line component by $0.3$, $0.5$, and $0.7$ and add it back to the unchanged underlying continuum to create three additional training continua. We renormalize each version over the same wavelength interval. Together with the 48 original continuum fits, these additional versions give a training set of 192 continua spanning a broader range of emission-line strengths.

We construct separate ten-component PCA bases over the full wavelength range and the red-side interval $1230$--$1599.2$\,\AA. Each basis describes the mean spectral shape and its principal variations across the training set. Working in logarithmic flux, the full-range reconstruction is
\begin{equation}
\ln \widehat C(\lambda)
=\mu_{\rm full}(\lambda)
+\sum_{k=1}^{10}a_k e_{k,\rm full}(\lambda).
\label{eq:continuum_pca}
\end{equation}
Here, $\widehat C(\lambda)$ is the normalized continuum, $\mu_{\rm full}(\lambda)$ is the mean logarithmic continuum, $e_{k,\rm full}(\lambda)$ is the $k$th PCA component, and $a_k$ is its coefficient for an individual quasar. The red-side basis has corresponding mean $\mu_{\rm red}(\lambda)$, components $e_{k,\rm red}(\lambda)$, and coefficients $b_k$.

We use the training continua to determine a linear relation between the red-side coefficients $b_k$ and the full-range coefficients $a_k$. To stabilize the prediction, we add a penalty proportional to the sum of the squared transformation coefficients, discouraging large values. This ridge penalty has strength $10^{-2}$; the fitted intercept is not penalized. For each DESI spectrum, we fit the red-side coefficients and use this relation to predict the full continuum in Equation~\ref{eq:continuum_pca}.

\subsection{Reconstructing DESI quasar continua}
\label{sec:continuum_desi}

Each DESI spectrum is shifted to the rest frame using the adopted quasar redshift and linearly resampled onto the common wavelength grid, propagating the pixel variances through the interpolation. We determine a normalization $\mathcal{N}$ from the inverse-variance-weighted mean flux over $1264$--$1296$\,\AA, rejecting outliers more than four robust standard deviations from the median. To reduce noise during the coefficient fit, we smooth the normalized spectrum with an inverse-variance-weighted Gaussian kernel of standard deviation 1.5 pixels and propagate the variances through this operation.

Let $\widetilde f_\lambda$ denote the smoothed normalized flux. We determine the red-side coefficients over $1230$--$1599.2$\,\AA\ by minimizing
\begin{equation}
\begin{aligned}
\chi^2_{\rm PCA}
&=\sum_{\lambda\in{\rm red}}w_\lambda
\Biggl[\ln \widetilde f_\lambda-\mu_{\rm red}(\lambda)
-\sum_{k=1}^{10}b_k e_{k,\rm red}(\lambda)\Biggr]^2
\\
&\quad+\eta_{\rm PCA}\sum_{k=1}^{10}b_k^2.
\end{aligned}
\label{eq:continuum_red_fit}
\end{equation}
The weights $w_\lambda$ are the squared signal-to-noise ratios of the smoothed pixels, capped at $10^4$ to prevent a few pixels from dominating the fit. The penalty strength is $\eta_{\rm PCA}=1$. Only pixels with finite positive flux and inverse variance enter the fit. We perform up to four iterations, rejecting pixels whose standardized residuals lie more than four robust standard deviations from their median. The fitted coefficients are then transformed into the full-range coefficients of Equation~\ref{eq:continuum_pca}.

We then smooth the predicted continuum and adjust the reconstructed \lya\ emission profile using red-side information. The prediction is smoothed blueward of $1200$\,\AA\ with a Gaussian kernel of standard deviation two pixels and blended linearly into the unsmoothed prediction between $1200$ and $1216$\,\AA. We shift the prediction to place its \lya\ emission maximum at $1215.67$\,\AA. To adjust the line shape, we multiply the prediction by $1+g_{\rm narrow}(\lambda)+g_{\rm broad}(\lambda)$, where the two Gaussian components have standard deviations of 2 and 8\,\AA. We fit their amplitudes to $f_{\rm obs}/C_{\rm pred}-1$ over $1216$--$1225$\,\AA, using inverse-variance weights and iterative outlier rejection. This interval lies redward of the proximity measurement. If the adjusted profile contains a local minimum between two maxima near \lya, we replace the intervening segment with a smooth interpolation that matches the surrounding values and slopes (a cubic Hermite interpolation). No measured flux blueward of \lya\ is used to determine these adjustments.

We reject reconstructions with fewer than 90 percent finite continuum pixels or a normalized continuum below 0.05 within $0$--$12{,}000\,\mathrm{km\,s^{-1}}$ blueward of \lya. We also require a valid normalization and at least 15 usable red-side pixels. The continuum-quality statistic used in Section~\ref{sec:data} is the weighted red-side residual sum, excluding the penalty, divided by the number of retained pixels minus the number of fitted coefficients. We use this statistic only to select acceptable continuum fits, since it neglects correlations introduced by resampling and smoothing.

Before the continuum corrections described in Sections~\ref{sec:meanflux}--\ref{sec:final_normalization}, the inferred transmission is
\begin{equation}
F_{\rm PCA}(\lambda)
=\frac{f_{\rm obs}(\lambda)}{\mathcal{N}\,\widehat C_{\rm PCA}(\lambda)},
\label{eq:continuum_transmission}
\end{equation}
where $f_{\rm obs}(\lambda)$ is the unsmoothed, resampled DESI flux and $\widehat C_{\rm PCA}(\lambda)$ is the normalized continuum after the \lya\ emission-line adjustments. Figure~\ref{fig:continuum_examples} illustrates the reconstructions for HST/COS spectra, mock spectra with known input continua, and DESI spectra.

\subsection{Continuum reconstruction validation}
\label{sec:continuum_tests}

We evaluate the red-to-blue PCA prediction using leave-one-out tests on the 48 HST/COS objects. For each object, we rebuild the PCA bases and the coefficient transformation using the remaining objects, excluding both the test object and all training continua derived from it. We fit its red-side spectrum and compare the predicted continuum with its empirical reference continuum.

We compare these predictions with estimates from LyCAN. LyCAN uses the spectrum over $1216$--$1600$\,\AA\ to predict the continuum over $1040$--$1600$\,\AA. We do not retrain the network for each test object. Because the HST/COS sample overlaps the data used to construct LyCAN's synthetic training continua, this comparison does not impose the same exclusion of test objects for both methods.

To compare the baseline performance of PCA and LyCAN near \lya, we evaluate their continuum predictions before applying our additional \lya\ emission-line adjustments. This separates the accuracy of the red-to-blue prediction from improvements introduced by those adjustments. The HST spectra, reference continua, and LyCAN predictions are each normalized by their own median over $1275$--$1285$\,\AA. The PCA training continua retain the $1264$--$1296$\,\AA\ normalization described above. We fit the red-side coefficients without inverse-variance weights, with a ridge penalty of $10^{-3}$ and up to four iterations rejecting residuals more than three sample standard deviations below their median. We smooth the predicted blue-side continuum as described above but omit the peak shift and the Gaussian and interpolation adjustments to the \lya\ emission profile used in the illustrative HST reconstructions in Figure~\ref{fig:continuum_examples}. The complete reconstruction procedure applied to DESI is tested separately using matched mock spectra.

For each object $i$, we define the fractional continuum residual
\begin{equation}
\delta_{C,i}(\lambda)
=\frac{C_{{\rm pred},i}(\lambda)-C_{{\rm ref},i}(\lambda)}
{C_{{\rm ref},i}(\lambda)}.
\label{eq:continuum_fractional_error}
\end{equation}
Figure~\ref{fig:continuum_method_comparison} shows the median residual across the HST/COS sample at each wavelength and the object-to-object scatter, measured as
\begin{equation}
\sigma_{\rm robust}(\lambda)
=\frac{P_{84}[\delta_C(\lambda)]-P_{16}[\delta_C(\lambda)]}{2},
\label{eq:continuum_robust_scatter}
\end{equation}
where $P_{16}$ and $P_{84}$ are the 16th and 84th percentiles. A systematic wavelength-dependent residual can distort the stacked transmission profile even as the random object-to-object errors average down.

\begin{table}[t]
\centering
\caption{HST/COS continuum-prediction errors over $0$--$10{,}000\,\mathrm{km\,s^{-1}}$ blueward of \lya. For each object, we average the absolute and signed fractional residuals over the valid wavelength pixels in this interval. Entries give the median of these averages across the HST/COS sample, expressed as percentages.}
\label{tab:continuum_validation}
\begin{tabular}{lrr}
\hline
Method & Absolute error (\%) & Signed error (\%) \\
\hline
PCA & 9.34 & $-0.79$ \\
LyCAN & 6.78 & $-1.98$ \\
\hline
\end{tabular}
\par\vspace{6pt}
\end{table}

Table~\ref{tab:continuum_validation} summarizes the errors over $0$--$10{,}000\,\mathrm{km\,s^{-1}}$ blueward of \lya. LyCAN has smaller typical absolute errors, while PCA has signed errors closer to zero. Both methods show their largest wavelength-dependent residuals near \lya\ (Figure~\ref{fig:continuum_method_comparison}). The PCA residual varies more smoothly there, although LyCAN has lower object-to-object scatter over much of the displayed range. We adopt PCA because its smaller signed residuals in these tests are desirable for measuring a stacked transmission profile.

The HST comparison tests the underlying continuum predictors. To assess the complete continuum-fitting procedure under conditions representative of our DESI sample, we use mock spectra that combine HST continuum shapes with \astrid{} absorption and DESI-like noise (Section~\ref{sec:astrid}). The known input continua allow us to measure and correct the systematic errors that remain after applying the same fitting procedure to the mock and DESI spectra (Section~\ref{sec:mock_continuum_correction}).

Before constructing the transmission stack, we further correct these continuum estimates using the external forest and matched mock spectra, as described in Sections~\ref{sec:meanflux}--\ref{sec:final_normalization}.

\section{Measuring the UV background from stacked transmission profiles}
\label{sec:method}

We first correct the reconstructed continua using the forest outside the proximity region and the known mock continua. We then stack the corrected transmission, fit the halo-model profiles, and calibrate the recovered UVBG using the matched mocks described in Section~\ref{sec:astrid}.

\subsection{External-forest mean-flux regulation}
\label{sec:meanflux}

Mean-flux regulation uses external measurements of the mean forest transmission to improve continuum estimates \citep{Lee2012,Lee2013}. We apply it to both mock and DESI spectra using the external forest at $1041$--$1160$\,\AA, entirely outside the proximity region. At rest-frame wavelength $\lambda$, the absorption redshift is $z_{\rm abs}=(1+z_q)\lambda/(1216\,\text{\AA})-1$, where $z_q$ is the quasar redshift. We use each DESI quasar's measured redshift and $z_q=3$ for the mock spectra. The expected transmission at each pixel is then $\langle F\rangle_{\rm ext}=\exp[-\bar\tau_{\rm eff}(z_{\rm abs})]$, using the \citet{Turner2024} relation in Equation~\ref{eq:mean_opacity_evolution}.

Following the anchored polynomial form of \citet{Lee2012}, we multiply each PCA continuum by
\begin{equation}
q(\lambda)=1+ax+cx^2,\qquad
x=\frac{\lambda-1280\,\text{\AA}}{100\,\text{\AA}}.
\label{eq:meanflux_correction}
\end{equation}
We determine $a$ and $c$ separately for each spectrum and set the multiplier to unity redward of $1280$\,\AA. We fit these coefficients within the external forest by comparing $F_{\rm PCA}$ with $q(\lambda)\langle F\rangle_{\rm ext}$. We weight the fit using the pixel-noise variance in transmission plus a fixed variance term of $0.2^2$. This additional term allows for fluctuations in forest absorption and prevents pixels with small measurement errors from dominating the continuum correction. We retain negative noisy flux. We require at least 30 usable pixels, a well-conditioned fit,\footnote{The smallest eigenvalue of the normal matrix must exceed $10^{-8}$ times its largest eigenvalue.} and a multiplier above 0.05 from the forest lower edge through $1216$\,\AA.

\subsection{Mock-derived wavelength-dependent continuum correction}
\label{sec:mock_continuum_correction}

After per-object regulation, we estimate the remaining systematic continuum residual as a function of wavelength across the matched \astrid{} mock sample, using the known input continua of the mock spectra constructed in Section~\ref{sec:mock_spectra}. We derive a common correction from these residuals and apply it to each individual mock and DESI continuum before stacking. For each mock, we calculate $\ln[C_{\rm true}/(q(\lambda)\,\mathcal N\,\widehat C_{\rm PCA})]$, where $q(\lambda)$ is the mean-flux correction defined in Equation~\ref{eq:meanflux_correction}. At each wavelength, we take the clipped, sightline-weighted mean of these logarithmic ratios, subtract its median over $1264$--$1296$\,\AA, smooth with a Gaussian of standard deviation two pixels, and renormalize in the same red-side interval. Exponentiating gives a common multiplier $M(\lambda)$, applied to both DESI and mock continua. We test the effect of this correction on UVBG recovery through the mock calibration in Section~\ref{sec:astrid_response}. We derive this correction entirely from the mock continua, without using the DESI transmission in the proximity region.

\subsection{Final external-forest normalization}
\label{sec:final_normalization}

After the per-object regulation and common wavelength correction, we combine the transmission spectra into a weighted mean stack as described in Section~\ref{sec:stack_construction}. We calculate a single normalization factor $s_{\rm F}$ from the external forest of this stack and divide the entire stacked transmission profile by it. We determine this factor separately for DESI and for the mock stack at each injected rate. It is the mean ratio of the measured to expected stacked transmission in the external forest,
\begin{equation}
s_{\rm F}=\left\langle
\frac{\overline F_{\rm corr}(\lambda)}
{\overline{\langle F\rangle}_{\rm ext}(\lambda)}
\right\rangle_{1041\text{--}1160\,\text{\AA}},
\label{eq:far_forest_slope}
\end{equation}
where the angle brackets denote an unweighted mean over valid wavelength pixels in the external forest and $\overline F_{\rm corr}$ is the stack after per-object regulation and the common wavelength correction. We use the same sightline weights and valid-pixel masks for the measured and expected stacks. Dividing the stacked transmission by this scalar is mathematically equivalent to multiplying every contributing continuum by the same factor, giving
\begin{equation}
C_{\rm final}(\lambda)=s_{\rm F}M(\lambda)q(\lambda)
\mathcal N\widehat C_{\rm PCA}(\lambda).
\label{eq:final_corrected_continuum}
\end{equation}
Here $\mathcal N$ restores the flux normalization defined in Section~\ref{sec:continuum_desi}. We test the effect of these corrections on UVBG recovery by applying the same correction sequence to the mocks. We obtain $s_{\rm F}=1.00725$ for DESI and 0.97799 for the mocks at injected $\Gamma_{12}=1$, where $\Gamma_{12}\equiv\Gamma_{\rm HI}/(10^{-12}\,\mathrm{s}^{-1})$. Sensitivity to the external mean opacity is discussed in Appendix~\ref{app:meanflux_sensitivity}.

\subsection{Assigning the DESI quasar bias}
\label{sec:desi_bias_assignment}

Our luminosity selection favors brighter quasars than the broader DESI population, so we adopt the four highest-luminosity bias measurements from the subsamples defined by fixed luminosity boundaries in Table~4 of \citet{Charles2026}.\footnote{The adopted bias anchors are $b_q=3.08\pm0.27$, $4.01\pm0.19$, $5.43\pm0.19$, and $6.03\pm0.31$ at $z=2.155$, $2.500$, $2.890$, and $3.279$, respectively. These are the statistical uncertainties used in the bias propagation.} To construct a continuous bias relation, we subtract their published redshift-only relation, $b_{\rm base}(z)=0.230[(1+z)^2-6.565]+2.394$, from each bias measurement, interpolate the offsets linearly in redshift, and add them back to $b_{\rm base}(z)$. We evaluate this relation for each DESI quasar. At $z=3$, it gives $b_q=5.593$.

We also test the profile-fit value $b_q=3.883$, obtained by fitting the halo model to the corrected, stacked \astrid{} mock transmission profile at a known injected UVBG. Section~\ref{sec:astrid_clustering} describes how we obtain this value and discusses its difference from the clustering bias. For this alternative DESI fit, we scale the bias relation down to match $b_q=3.883$ at $z=3$, retaining its redshift dependence. Both bias choices yield similar calibrated UVBG measurements, as shown in Section~\ref{sec:combined_uvbg}.

\subsection{Constructing and fitting transmitted flux profiles}
\label{sec:stack_construction}

We infer $\GammaHI$ by fitting stacked theoretical halo-model transmission profiles to the observed DESI transmission stack. We evaluate 321 candidate values of $\GammaHI$, logarithmically spaced between $10^{-14}$ and $10^{-10}\,\mathrm{s}^{-1}$. For each candidate value, we construct a corresponding theory stack using the measured luminosities and redshifts of the DESI quasars, the quasar bias $b_q$ assigned in Section~\ref{sec:desi_bias_assignment}, and the same signal-to-noise-based weights used to construct the observed stack. We then compare each candidate theory stack with the measured DESI stack to determine which value of $\GammaHI$ is best supported by the data.

To construct a theory stack, we first predict a transmission curve for each quasar. For a specified candidate $\GammaHI$, the measured quasar luminosity determines the equality radius $r_{\rm eq}$, at which the photoionization rate from the quasar equals that of the UVBG. The adopted quasar bias $b_q$ determines the environmental contribution, while the quasar redshift sets the appropriate model epoch and opacity evolution. Consequently, quasars with different luminosities and redshifts contribute different theory curves even when testing the same value of $\GammaHI$.

Rather than recomputing the halo model for every quasar and candidate $\GammaHI$, we use a precomputed grid of transmission profiles spanning redshift, equality radius, and quasar bias. Before interpolating this grid, we apply the $r_{\rm eq}$ remapping described in Appendix~\ref{app:radius_mapping} to both the mock and DESI fits. We interpolate linearly in redshift and in the logarithms of $r_{\rm eq}$ and $b_q$. Finally, we rescale the model optical depth across the profile using the mean-opacity evolution defined in Equations~\ref{eq:mean_opacity_evolution} and \ref{eq:profile_redshift_evolution} of Section~\ref{sec:model_extensions}. We then interpolate the transmission onto the spectral wavelength grid and smooth it with a Gaussian of standard deviation $500\kms$, renormalizing the kernel at the boundaries of the supported profile. The resulting curve gives the predicted transmission as a function of velocity separation for that quasar.

At each wavelength pixel $j$, we calculate the weighted mean transmission across quasars,
\begin{equation}
\overline F_j =
\frac{\sum_q w_q u_{qj}F_{qj}}
{\sum_q w_q u_{qj}},
\end{equation}
where $w_q$ is the weight assigned to quasar $q$, and $u_{qj}=1$ for retained pixels and zero otherwise. The quasar weights $w_q$ are the square of the red-side signal-to-noise statistic in Equation~\ref{eq:snr}, capped at $S/N=10$. This cap prevents a small number of high-S/N quasars from dominating either stack. We then bin the stacked transmission profile into radial bins defined by line-of-sight velocity separation from the quasar. The bin boundaries are 200, 1200, 2200, 3200, 4500, 6500, 9000, and $9900\kms$. The outermost bin is truncated at the last sampled point of the halo-model profiles. The DESI and mock stacks use the same seven bins, with lower edges included and upper edges excluded. Within each bin $i$, we average the stacked transmission over its wavelength pixels,
\begin{equation}
d_i=\frac{1}{n_i}\sum_{j\in i}\overline F_j,
\label{eq:transmission_stack}
\end{equation}
where the sum includes the $n_i$ wavelength pixels with valid stacked transmission. At each wavelength, we compare transmissions across the quasar sample and reject values more than three robust standard deviations from the median, using 1.4826 times the median absolute deviation to estimate the scatter. The resulting clipping mask identifies which available pixels are retained in the stack. The full-sample data clipping masks are held fixed during bootstrap resampling.

We construct the theory stack with the same order of wavelength stacking and radial averaging, using the quasars' measured luminosities and redshifts and the adopted $b_q$ at each redshift. The theory follows the finite-pixel coverage of the spectra, but its clipping mask is calculated from the predicted transmissions rather than copied from the observed stack.

We denote the predicted mean transmission in radial bin $i$ by $m_i(\GammaHI)$. For each candidate photoionization rate, we compare the model vector $\boldsymbol{m}(\GammaHI)$ with the measured transmission vector $\boldsymbol{d}$. We define their difference as $\boldsymbol{\Delta}=\boldsymbol{d}-\boldsymbol{m}(\GammaHI)$ and calculate
\begin{equation}
\chi^2(\GammaHI)
=
\boldsymbol{\Delta}^{\mathsf T}
\mathbf C^{-1}
\boldsymbol{\Delta}.
\label{eq:profile_likelihood}
\end{equation}
Here $\mathbf C$ is the covariance matrix of the measured stack, and $\mathbf C^{-1}$ denotes its estimated inverse with the finite-sample correction described below.

The candidate photoionization rates are spaced by 0.0125 dex in $\log_{10}\GammaHI$. To refine the best-fitting rate between these sampled values, we fit a quadratic to $\chi^2$ as a function of $\log_{10}\GammaHI$, using the lowest sampled $\chi^2$ and its two adjacent points. We determine the conditional 68 percent fit interval from the rates on either side of the minimum where $\chi^2$ rises by one, using linear interpolation between sampled points in $\log_{10}\GammaHI$. Candidate rates that require predictions outside the available halo-model grid are excluded.

Each fit uses seven radial bins and has one free parameter, $\GammaHI$, giving six degrees of freedom.

We estimate the covariance of each stack from 512 bootstrap realizations, resampling whole quasar sightlines, and denote the resulting sample covariance by $\mathbf S$. To reduce noise in the estimated correlations between radial bins and stabilize the covariance inversion, we adopt 30 percent diagonal shrinkage. This leaves the diagonal entries unchanged and reduces the off-diagonal entries by 30 percent. We also add an uncertainty $\sigma_{\rm mod}$ in quadrature to each radial bin to allow for discrepancies between the model and measured transmission. The covariance used in the fit is
\begin{equation}
\mathbf C=0.7\mathbf S+0.3\operatorname{diag}(\mathbf S)
+\sigma_{\rm mod}^2\mathbf I,
\end{equation}
where $\operatorname{diag}(\mathbf S)$ retains only the diagonal entries of $\mathbf S$, and $\mathbf I$ is the identity matrix.  We adopt $\sigma_{\rm mod}=0.015$ for both DESI and \astrid{}, in absolute transmission units. This model-error floor allows for model inaccuracies and remaining observational systematics without introducing correlations between radial bins. We explain the adopted values in Section~\ref{sec:model_error}.

Estimating a covariance matrix from a finite number of realizations can bias its inverse. We apply the correction of \citet{Hartlap2007}, multiplying the pseudoinverse of $\mathbf C$ by $(N_{\rm boot}-N_{\rm bin}-2)/(N_{\rm boot}-1)=503/511\simeq0.984$ for our 512 bootstrap realizations and seven radial bins. We denote the corrected inverse by $\mathbf C^{-1}$ in Equation~\ref{eq:profile_likelihood}. The correction is approximate for our covariance, which also includes diagonal shrinkage and a model-error floor.

\section{Validation and calibration with \astrid{} mock spectra}
\label{sec:astrid}

We construct mock quasar spectra using the \astrid{} hydrodynamic simulation to test how accurately our analysis recovers a known ultraviolet background. We assign each simulated quasar a luminosity drawn from our DESI sample and generate transmission profiles for a series of injected photoionization rates, $\GammaHI^{\rm inj}$. We combine these profiles with HST continuum shapes and DESI-like noise, then apply the continuum reconstruction and proximity-profile fitting described in Sections~\ref{sec:continuum} and \ref{sec:method}. Comparing the recovered photoionization rates with their known input values allows us to calibrate systematic offsets in the inferred UVBG.

\subsection{The \astrid{} hydrodynamic simulation}
\label{sec:astrid_simulation}

We use the \astrid{} cosmological hydrodynamic simulation to model the gas environments around quasars. The simulation evolves gas density, temperature, and velocity under gravity and hydrodynamics, including cooling and feedback. Its absorption profiles therefore provide a more detailed treatment of these environments than the analytic halo model.

\astrid{} was performed with an updated version of the \texttt{MP-Gadget} code \citep{Bird2022,Croft2025}. We use the $z=3$ snapshot studied by \citet{Croft2025}. The periodic simulation volume is a cube $250\,h^{-1}\,\mathrm{cMpc}$ on a side and initially contains $5500^3$ dark-matter particles and the same number of smoothed-particle-hydrodynamics gas elements. The dark-matter and initial gas-particle masses are $6.74\times10^6\,h^{-1}\,M_\odot$ and $1.27\times10^6\,h^{-1}\,M_\odot$, respectively, and the gravitational softening length is $1.5\,h^{-1}\,\mathrm{kpc}$. The simulation cosmology has $\Omega_{\rm m}=0.3089$, $\Omega_\Lambda=0.6911$, $\Omega_{\rm b}=0.0486$, $h=0.6774$, $\sigma_8=0.82$, and $n_{\rm s}=0.9667$.

The simulation includes radiative and metal-line cooling, star formation, stellar feedback, black-hole seeding and growth, and thermal active-galactic-nucleus feedback. At $z=3$, the simulated intergalactic medium is highly ionized. For mock-spectrum construction, we use a spatially uniform ultraviolet radiation background and add the ionizing radiation from each quasar during spectral post-processing.

We use the sightline extraction and absorption-spectrum calculations described by \citet{Croft2025}, which we summarize here. The \lya\ spectra begin at simulation quasar positions and extend parallel to a box axis. Neutral-hydrogen density, temperature, and line-of-sight peculiar velocity are projected from the gas elements onto spectral pixels. We reduce the optical depth to account for the quasar radiation and map the spectra into redshift space, including peculiar velocities and thermal broadening. The resulting profiles retain the simulated variations in gas density, velocity, and temperature around quasars.

\subsection{Constructing mock DESI spectra}
\label{sec:astrid_hybrid_mocks}

We begin with 1,000 unique \astrid{} sightlines and pair each with a quasar from our DESI sample. The measured DESI redshift sets the conversion between rest-frame and observed wavelength, while the measured luminosity determines the ionizing contribution assigned to the simulated quasar. This reassignment is necessary because the \astrid{} volume does not contain enough quasars as luminous as those selected from the much larger DESI survey volume. Following \citet{Croft2025}, we reassign the luminosities while preserving the gas density, velocity, and temperature structure of each sightline. The initial matching targets the redshift, luminosity, and red-side signal-to-noise distributions of the DESI sample. Of the initial 1,000 mocks, \ZThreeNASTRID{} pass the selection described below and are retained for calibration.

For each assigned luminosity, we vary the injected UVBG through the equality radius, $r_{\rm eq}\propto(\GammaHI^{\rm inj})^{-1/2}$. The quasar contribution reduces the real-space optical depth by the factor $[1+\Gamma_{\rm QSO}(r)/\GammaHI^{\rm inj}]^{-1}$ before the velocity and thermal broadening are applied. Varying the injected rate therefore changes the contrast between quasar radiation and the background while retaining the underlying forest absorption. We generate seventeen input rates spanning $0.25\leq\Gamma_{12}^{\rm inj}\leq32$.

The native \astrid{} profiles contain 50 transmission bins spanning $100$--$9900\kms$ blueward of \lya. Before interpolating these profiles onto the spectral wavelength grid, we apply the mean-opacity evolution described in Section~\ref{sec:model_extensions},
\begin{equation}
F_{\rm evolved}(v)=F_{\rm snapshot}(v)^{\bar\tau_{\rm eff}(z_{\rm abs})/\bar\tau_{\rm eff}(3)},
\end{equation}
where $1+z_{\rm abs}=4(1-v/c)$. This accounts for the changing mean absorption along a sightline from a quasar at $z=3$, without evolving the simulated gas density, temperature, or velocity. The assigned DESI redshift sets only the observed wavelengths of the mock spectrum. We apply the opacity evolution once, before adding noise or fitting the continuum, and use the same evolution in the halo-model predictions.

At wavelengths blueward of the native proximity profiles, we append a correlated lognormal forest realization to supply the external forest used for continuum correction. Its target mean transmission is $\exp[-\bar\tau_{\rm eff}(z_{\rm abs})]$, using the relation in Equation~\ref{eq:mean_opacity_evolution}. Gaussian fluctuations in logarithmic transmission are smoothed with a kernel of width $200\kms$ and normalized to a standard deviation of 0.5. After exponentiation, the transmission is restricted to $0\leq F\leq1$, with its mean adjusted to reproduce the target value. This extension already includes opacity evolution and receives no further correction. We keep the same external-forest realization at every injected UVBG.

\subsubsection{Mock spectra, selection, and continuum fitting}
\label{sec:mock_spectra}

QuickQuasars \citep{HerreraAlcantar2025} supplies the flux normalization, inverse variance, and missing-pixel mask for each matched object. We replace its intrinsic continuum with an HST reference continuum from Section~\ref{sec:continuum_training}. We use templates that are finite across the adopted wavelength grid and have positive median flux over $1264$--$1296$\,\AA, distributing them approximately equally among the mock objects. Each template is normalized over this interval and scaled to the outlier-clipped, inverse-variance-weighted mean flux of the matched QuickQuasars spectrum.

For each mock spectrum at injected photoionization rate $\GammaHI^{\rm inj}$, the flux is
\begin{equation}
f(\lambda)=A C(\lambda)T(\lambda,\GammaHI^{\rm inj})+n(\lambda).
\end{equation}
Here $A$ is the flux normalization, $C$ is the assigned HST continuum normalized to unit median flux over $1264$--$1296$\,\AA, and $T$ is the simulated transmission. We draw Gaussian pixel noise $n$ with variance given by the reciprocal of the QuickQuasars inverse variance and retain its missing-pixel mask. We use one noise realization for the main calibration and a second independent realization to test sensitivity to noise (Appendix~\ref{app:mock_noise}). Within each realization, the assigned continuum, pixel noise, and observational properties are held fixed across injected rates.

We reconstruct the mock continua using the procedure in Section~\ref{sec:continuum}. For the main calibration, the PCA training set excludes the HST quasar whose continuum was used to construct that mock, including its augmented versions. This prevents the predictor from being trained on the continuum it is asked to recover. We then apply the external-forest correction, the common mock-derived wavelength correction, and the final normalization described in Sections~\ref{sec:meanflux}--\ref{sec:final_normalization}. These mocks provide the known input continua used to derive the wavelength-dependent continuum correction in Section~\ref{sec:mock_continuum_correction}.

Each mock must have red-side $S/N\geq3$ as defined in Equation~\ref{eq:snr}, a continuum-fit quality statistic no larger than 5 as defined in Section~\ref{sec:continuum_desi}, and at least 85 percent coverage of the fitted region. The assigned luminosities already satisfy $M_{1450}\leq-25.5$. We require successful continuum fits with both the full and leave-one-out training sets in both noise realizations, as well as successful external-forest correction across the injected rates. The final sample contains \ZThreeNASTRID{} mock quasars representing \ZThreeNTemplates{} HST templates, used at all injected rates.

\begin{figure*}[t]
\centering
\zfigure[0.95\textwidth]{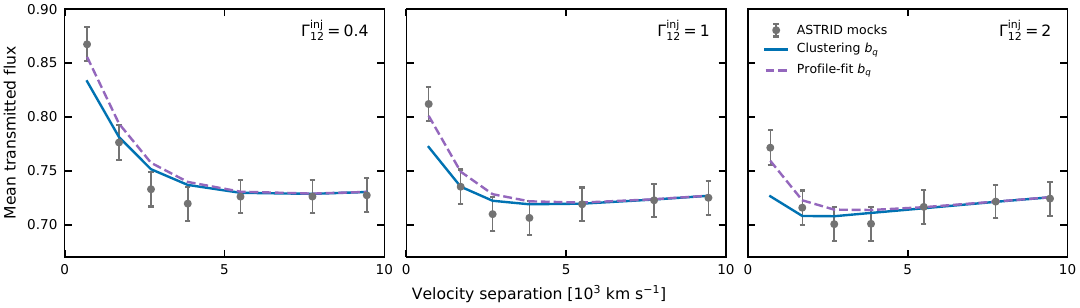}
\caption{Stacked \astrid{} mock transmission (gray points) after applying the continuum corrections described in Sections~\ref{sec:meanflux}--\ref{sec:final_normalization}, compared with halo-model predictions at injected rates $\Gamma_{12}^{\rm inj}=0.4$, 1, and 2. In each panel, both model curves use the indicated injected rate. Solid blue curves use the clustering-based bias $b_q=5.593$ (Section~\ref{sec:astrid_clustering}). Dashed purple curves use $b_q=\ZThreeEffectiveBias$, obtained by fitting the halo model to the mock transmission profile at $\Gamma_{12}^{\rm inj}=1$ (Section~\ref{sec:astrid_clustering}). Points show the seven fitted radial bins, and lines join the model bin averages. Error bars include the bootstrap covariance and the 0.015 transmission model-error floor.}
\label{fig:astrid_shape_comparison}
\end{figure*}

\subsubsection{Quasar bias of the \astrid{} mocks}
\label{sec:astrid_clustering}

For DESI, we adopt the quasar bias inferred from the clustering measurements of \citet{Charles2026}, as described in Section~\ref{sec:desi_bias_assignment}. To determine the corresponding $b_q$ for our mock sample, we measure the clustering of the selected \astrid{} quasars using their three-dimensional positions. We count quasar pairs as a function of separation, weighting each pair by $w_iw_j$, where $w_i$ are the same red-side signal-to-noise weights used for the mock calibration. Pair separations account for the periodic boundaries of the simulation box.

We calculate the quasar correlation function as $\xi_{qq}(r)=DD/RR-1$, where $DD$ is the weighted pair count in a separation bin and $RR$ is the expected count for objects distributed uniformly throughout the box. For a separation bin with shell volume $V$,
\begin{equation}
RR=\frac{V}{L^3}\frac{(\sum_iw_i)^2-\sum_iw_i^2}{2},
\end{equation}
where $L=250\,h^{-1}\,\mathrm{cMpc}$. We determine $b_q$ by fitting $\xi_{qq}(r)=b_q^2\xi_{mm}(r)$ over $10$--$36\,h^{-1}\,\mathrm{cMpc}$, using a linear-theory matter correlation function $\xi_{mm}$ averaged over the same separation bins.

We obtain $b_q=\ZThreeCatalogueBias\ZThreeCatalogueBiasError$. We estimate the uncertainty by dividing the simulation volume into 27 regions and repeating the measurement with each region omitted in turn. A uniform random catalog accounts for the missing region when calculating the expected pair counts. This measurement agrees with the bias adopted for DESI from \citet{Charles2026}, $b_q(z=3)=5.593$, well within its uncertainty. We therefore use $b_q=5.593$ for the primary mock calibration as well as the corresponding redshift-dependent relation for DESI.

In addition to the clustering bias, we test the value of $b_q$ for which the model best matches the corrected mock transmission profile at injected $\Gamma_{12}=1$. We hold the radiation contribution fixed and minimize $\chi^2$ over $1\leq b_q\leq8$ using all seven radial bins. This gives $b_q=\ZThreeEffectiveBias$, labeled \textit{Profile-fit} in the figures. We hold this value fixed when fitting mocks at other injected rates and when propagating calibration uncertainty. Each bias choice has its own UVBG calibration.

The lower profile-fit bias may reflect differences in the gas temperature and density distribution that are not fully captured by the halo model. Simulations show that these properties affect Ly$\alpha$ absorption around quasar hosts \citep{Sorini2018}, and that feedback can increase transmission at fixed matter overdensity \citep{Dong2024}. Such an effect could cause the halo model to favor a lower $b_q$ to reproduce the mock absorption, even when the quasar clustering implies a higher value. Regardless of the origin of this discrepancy, we repeat the analysis with both bias values and obtain consistent calibrated UVBG measurements, as shown in Section~\ref{sec:combined_uvbg}.

\begin{figure*}[t]
\centering
\zfigure[\textwidth]{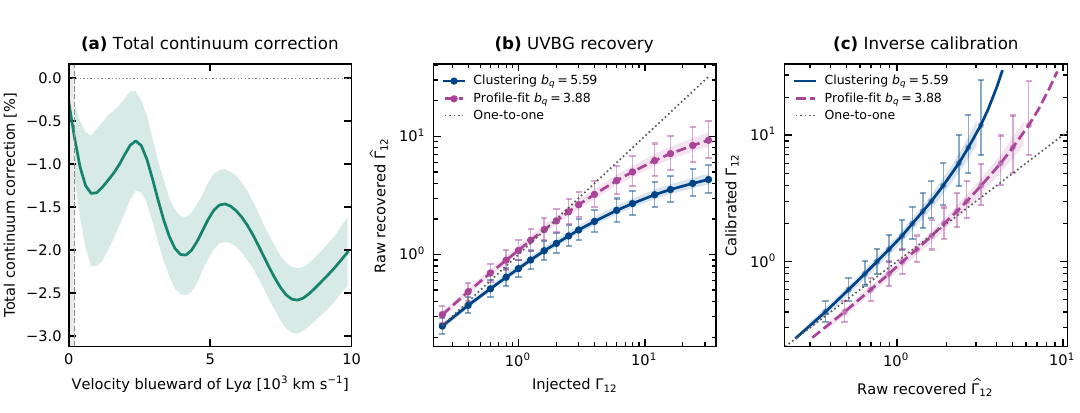}
\caption{Continuum correction and UVBG calibration using the matched \astrid{} mocks. \textbf{(a)} Overall continuum correction at injected $\Gamma_{12}=1$, averaged over mock sightlines using the stacking weights and expressed as a percentage change from the PCA continuum. It combines the external-forest correction, common wavelength correction, and final normalization described in Sections~\ref{sec:meanflux}--\ref{sec:final_normalization}. Shading shows the 16th--84th percentile interval from bootstrap realizations. \textbf{(b)} Photoionization rate recovered after applying the complete continuum reconstruction, correction, and halo-model fitting procedure to mock spectra, plotted against the known injected rate. \textbf{(c)} The inverse of this relation, used to convert a fitted photoionization rate into a calibrated UVBG estimate. Blue and purple denote the clustering and profile-fit bias choices. Shaded bands in panels (b) and (c) show bootstrap uncertainty in the response and its inverse. Error bars in panel (b) show the $\Delta\chi^2=1$ intervals from the mock profile fits, with the continuum corrections and quasar bias held fixed (Section~\ref{sec:method}). In panel (c), these intervals are converted using the calibration curve derived from the full mock sample, with that curve held fixed. Dotted lines mark one-to-one recovery.}
\label{fig:response}
\end{figure*}

\subsection{Calibrating the recovered UV background}
\label{sec:astrid_response}
\label{sec:astrid_shape_comparison}

We now apply the proximity-effect analysis to the mock spectra to test how accurately the halo model recovers the injected UVBG. We first compare the model and mock transmission profiles at the same photoionization rate, then fit the mock profiles and use the recovered rates to calibrate the DESI measurement.

Figure~\ref{fig:astrid_shape_comparison} compares the stacked mock transmission profiles with halo-model predictions at three injected photoionization rates. The mock spectra have undergone the continuum reconstruction and corrections described in Section~\ref{sec:continuum} and Sections~\ref{sec:meanflux}--\ref{sec:final_normalization}. In each panel, the model uses the same UVBG as the mocks, and both include the mean-opacity evolution described in Section~\ref{sec:model_extensions}. Using the profile-fit bias improves agreement in the innermost radial bin while retaining agreement within the plotted uncertainties at larger separations. Across the three injected rates, the halo model reproduces the overall radial dependence of the mock transmission.

Figure~\ref{fig:response}(a) shows that the overall continuum correction lowers the PCA continuum by roughly 1--3 percent over most of the fitted region, raising the inferred transmission. The corrections described in Sections~\ref{sec:meanflux}--\ref{sec:final_normalization} adjust both the level and radial shape of the transmission profile.

For each injected rate, we reconstruct and correct the mock continua, stack the transmission, and fit for the photoionization rate using the procedure in Section~\ref{sec:method}. Figure~\ref{fig:response}(b) shows the recovered rates as a function of the injected values. We denote the recovered rate by $\widehat\Gamma_{12}$ and write this relation as $\widehat\Gamma_{12}=R(\Gamma_{12}^{\rm inj})$. We invert it to convert the fitted DESI rate into a calibrated estimate,
\begin{equation}
\Gamma_{12}^{\rm cal}=R^{-1}(\widehat\Gamma_{12}),
\label{eq:response_inverse}
\end{equation}
as illustrated in Figure~\ref{fig:response}(c). We check that the recovered photoionization rate increases monotonically with the injected rate, then invert this relation using linear interpolation in logarithmic space.

Figure~\ref{fig:response}(b) shows a smooth, monotonic response, with the recovered rate approximately proportional to the injected rate around $\Gamma_{12}^{\rm inj}\sim1$. At large injected UVBG values, the response flattens and the recovered rate falls below the input. The inverse calibration in panel (c) compensates for this change in response, allowing us to convert fitted rates into calibrated UVBG estimates.

For the primary DESI measurement, the calibration changes the fitted rate from $\widehat\Gamma_{12}=\ZThreeClusteringRaw$ to $\Gamma_{12}=\ZThreeClusteringGamma$. Using the profile-fit bias changes the corresponding estimate from $\ZThreeEffectiveRaw$ to $\ZThreeEffectiveGamma$. The calibrated measurements are similar despite the different fitted rates before calibration. This agreement suggests that the inferred UVBG responds similarly to changes in the assumed $b_q$ in DESI and the \astrid{} mocks, supporting the use of the mocks to calibrate the DESI measurement (Section~\ref{sec:discussion}). 

The mock calibration depends on how well the HST continuum shapes represent the DESI quasars. We test its transfer between the available continuum shapes by deriving the calibration from mocks assigned one group of HST templates and applying it to mocks assigned a disjoint group (Appendix~\ref{app:template_groups}). We also test an independent pixel-noise realization (Appendix~\ref{app:mock_noise}). These tests use the covariance matrix described in Section~\ref{sec:stack_construction}. Sensitivity to the mock model-error floor is examined in Appendix~\ref{sec:floor_sensitivity}.

\section{Results}
\label{sec:results}

\subsection{Halo-model fits to DESI observations}

\begin{figure}[t]
\centering
\zfigure[\columnwidth]{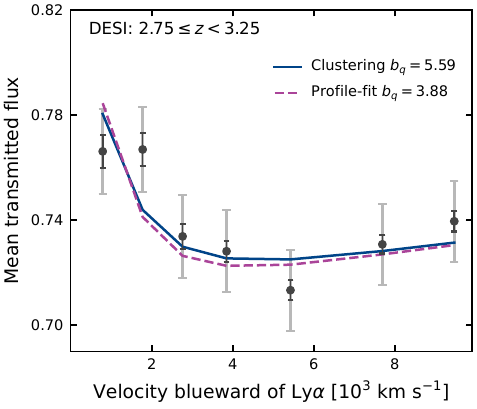}
\caption{Stacked DESI transmission profile and best-fitting halo-model predictions. Gray points show the measured transmission in the seven radial bins after continuum correction. Dark inner bars show sightline-bootstrap errors; pale outer bars also include the model-error floor of 0.015 in transmission (Section~\ref{sec:model_error}). Solid blue and dashed purple curves show fits using the clustering and profile-fit biases, respectively.}

\label{fig:profiles}
\end{figure}

We now combine the preceding steps to measure the UVBG from the DESI sample selected in Section~\ref{sec:data}. Using the continua reconstructed in Section~\ref{sec:continuum}, we apply the continuum corrections, quasar-bias assignments, and stacking and fitting procedure described in Section~\ref{sec:method} to fit the halo model of Section~\ref{sec:model} to the measured transmission. We then convert the fitted photoionization rates into calibrated UVBG estimates using the mock-derived calibration in Section~\ref{sec:astrid_response}. Figure~\ref{fig:profiles} shows the resulting stacked transmission profile and its halo-model fits. Both bias choices described in Section~\ref{sec:desi_bias_assignment} reproduce the broad decline in transmission away from the quasar and the shallow rise at larger separations, yielding nearly overlapping fitted profiles. Both provide good fits with the adopted uncertainties, with reduced $\chi^2$ values of \ZThreeClusteringChi{} and \ZThreeEffectiveChi{} for the clustering and profile-fit biases, respectively, for six degrees of freedom.

\begin{figure*}[t]
\centering

\zfigure[\textwidth]{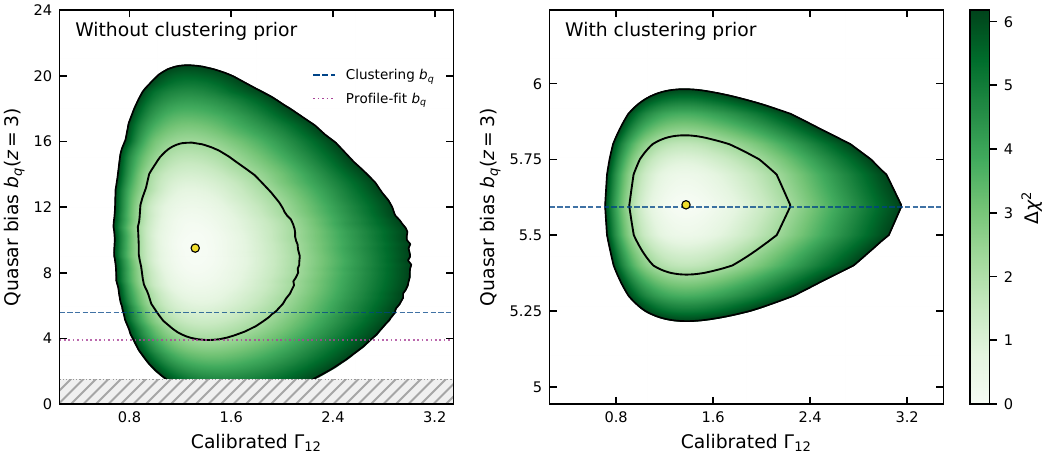}
\caption{Joint constraints on calibrated $\Gamma_{12}$ and quasar bias $b_q(z=3)$ from the DESI transmission profile. The left panel shows $\Delta\chi^2$ from fits of the halo model to the DESI transmission profile at different trial UVBG values and quasar biases. The right panel adds the Gaussian clustering-prior term $[(b_q-b_{\rm clust})/\sigma_b]^2$ to the transmission-fit $\chi^2$, where $b_{\rm clust}$ and $\sigma_b$ are the clustering estimate of quasar bias at $z=3$ and its statistical uncertainty. Contours mark $\Delta\chi^2=2.30$ and $6.18$ relative to the minimum in each panel, corresponding to nominal 68.3 and 95.4 percent joint confidence regions for two parameters. Blue dashed and purple dotted horizontal lines mark the clustering and profile-fit biases. The horizontal axis shows the photoionization rate after applying the mock calibration for each assumed quasar bias. The hatched region lies outside the supported calibration range. Fits follow Section~\ref{sec:stack_construction} and include the DESI model-error floor of 0.015, but omit uncertainty in the continuum correction and calibration and the additional systematic contributions in Section~\ref{sec:systematics}.}

\label{fig:uvb_bq_contours}
\end{figure*}

\begin{table*}[!t]
\centering
\caption{UVBG measurements from our DESI DR1 quasar sample described in Section~\ref{sec:sample_selection}. Statistical errors are distances to the DESI-bootstrap 16th and 84th percentiles. The systematic uncertainty includes the contributions in Section~\ref{sec:expanded_budget}. The reduced $\chi^2$ values describe the transmission-profile fits before calibration, include the DESI model-error floor of 0.015, and use six degrees of freedom.}
\label{tab:allz}
\small
\begin{tabular}{@{}lccccc@{}}
\toprule
Bias choice & $b_q(z=3)$ & $\Gamma_{12}^{\rm cal}$ & Statistical error & Systematic uncertainty & $\chi^2/6$\\
\midrule
Clustering & \ZThreeClusteringBias & \ZThreeClusteringGamma & $^{+\ZThreeClusteringSamplingPlus}_{-\ZThreeClusteringSamplingMinus}$ & $^{+\ZThreeClusteringSystematicPlus}_{-\ZThreeClusteringSystematicMinus}$ & \ZThreeClusteringChi\\
Profile-fit & \ZThreeEffectiveBias & \ZThreeEffectiveGamma & $^{+\ZThreeEffectiveSamplingPlus}_{-\ZThreeEffectiveSamplingMinus}$ & $^{+\ZThreeEffectiveSystematicPlus}_{-\ZThreeEffectiveSystematicMinus}$ & \ZThreeEffectiveChi\\
\bottomrule
\end{tabular}
\end{table*}

\subsection{The UVBG measurement at $z\simeq3$}
\label{sec:combined_uvbg}

Applying the mock calibration in Equation~\ref{eq:response_inverse} to the fit using the quasar bias $b_q(z)$ inferred from clustering gives our primary measurement,
\begin{equation}
\boxed{\GammaHI=\left(\ZThreeClusteringGamma^{+\ZThreeClusteringSamplingPlus}_{-\ZThreeClusteringSamplingMinus}\right)\times10^{-12}\,\mathrm{s}^{-1}}
\label{eq:headline}
\end{equation}
at $z_{\rm eff}=\ZThreeZEffective$. This effective redshift is the mean quasar redshift weighted in the same way as the transmission stack. The quoted statistical errors are estimated by bootstrap resampling the full set of DESI sightlines. The systematic uncertainty is $^{+\ZThreeClusteringSystematicPlus}_{-\ZThreeClusteringSystematicMinus}\times10^{-12}\,\mathrm{s}^{-1}$. Section~\ref{sec:expanded_budget} describes how we calculate the statistical and systematic uncertainties.

The statistical errors give the distances from the central estimate of the UVBG photoionization rate to the 16th and 84th percentiles of its bootstrap distribution. The systematic uncertainty estimates the effects of uncertainties in our modeling assumptions, continuum corrections, calibration, and external inputs, as described in Section~\ref{sec:systematics}. Repeating the analysis with the profile-fit bias derived in Section~\ref{sec:astrid_clustering} gives $\Gamma_{12}=\ZThreeEffectiveGamma$. The two calibrated central estimates differ by only 3.3 percent. The upper and lower errors differ because the relation between transmission and photoionization rate, including the mock calibration, is nonlinear. Table~\ref{tab:allz} summarizes both results.

\subsection{Degeneracy between the UVBG and quasar bias}
\label{sec:uvb_bias_degeneracy}

Figure~\ref{fig:uvb_bq_contours} shows the joint constraints on quasar bias and calibrated UVBG. The left panel shows $\Delta\chi^2$ from the transmission-profile fits alone. The broad contours show that the transmission profile does not precisely determine both the quasar bias and the UVBG simultaneously. We therefore use an external clustering constraint to determine the quasar bias independently. We adopt the clustering measurements of \citet{Charles2026} as described in Section~\ref{sec:desi_bias_assignment}. The right panel includes this constraint as a Gaussian prior on $b_q(z=3)$, giving
\begin{equation}
\chi^2_{\rm tot}(\Gamma_{12},b_q)
=\chi^2_{\rm trans}(\Gamma_{12},b_q)
+\left(\frac{b_q-b_{\rm clust}}{\sigma_b}\right)^2,
\label{eq:clustering_prior_chi2}
\end{equation}
where $\chi^2_{\rm tot}$ is the combined statistic, $\chi^2_{\rm trans}$ is the transmission-fit statistic, and $b_{\rm clust}$ and $\sigma_b$ are the clustering estimate of quasar bias at $z=3$ and its statistical uncertainty. We subtract the minimum of $\chi^2_{\rm trans}$ in the left panel and the minimum of $\chi^2_{\rm tot}$ in the right panel to obtain the plotted $\Delta\chi^2$ values.
Comparing the two panels shows that adding the clustering prior substantially narrows the allowed range of $b_q$ around $b_q(z=3)=5.6$, while producing only a small shift in the best-fitting calibrated UVBG. At each trial bias, we apply the corresponding mock-derived calibration and hold it and the continuum correction fixed when calculating the contours. The contours do not include the additional systematic uncertainties discussed in Section~\ref{sec:expanded_budget}.

\begin{figure*}[!t]
\centering
\zfigure[0.90\textwidth]{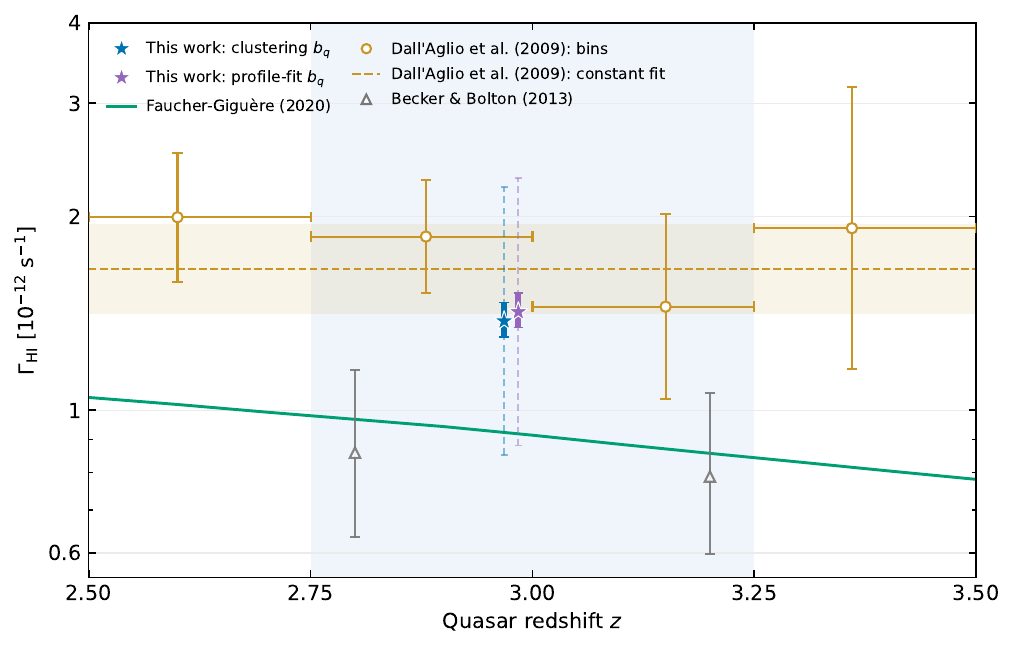}
\caption{Hydrogen photoionization rate versus redshift. Blue and purple filled stars show the calibrated estimates using the clustering and profile-fit quasar biases, respectively. Both measurements are for the same sample at $z\simeq3$; their horizontal separation is purely cosmetic. Solid inner error bars show statistical uncertainties estimated from the 16th and 84th percentiles of bootstrap resamples of the DESI sightlines. These can be compared with the bootstrap error bars of \citet{DallAglio2009}. Pale dashed outer error bars combine statistical uncertainty with the systematic uncertainty in Section~\ref{sec:expanded_budget}, added in quadrature in the logarithmic rate. The pale-blue window marks the sample range $2.75\leq z<3.25$. Gold open circles show the binned modal estimates of \citet{DallAglio2009}, with redshifts and bin boundaries from their Table~1 and bootstrap errors digitized from the solid-circle series in their Figure~9. Horizontal bars mark the bin boundaries. The gold dashed line and band show their constant fit and quoted uncertainty over their full redshift sample. Gray triangles show \citet{BeckerBolton2013}, and the green curve shows the model of \citet{FaucherGiguere2020}.}

\label{fig:evolution}
\end{figure*}

\subsection{Comparison with previous measurements}

Figure~\ref{fig:evolution} compares our calibrated $z\simeq3$ measurement with previous measurements and a UVBG model. We compare with the line-of-sight proximity-effect measurements of \citet{DallAglio2009} and the mean Ly$\alpha$ forest opacity measurements interpreted using hydrodynamical simulations by \citet{BeckerBolton2013}. We also show the observationally constrained UVBG model of \citet{FaucherGiguere2020}, which combines radiation from galaxies and active galactic nuclei with intergalactic absorption. Our estimate is approximately 17 percent below the average level reported by \citet{DallAglio2009}, with overlapping uncertainties. A lower estimate is consistent with the expected direction of accounting explicitly for the quasar environment. Enhanced environmental absorption can otherwise be mistaken for a stronger UVBG \citep{LoebEisenstein1995,FaucherGiguere2008}. Our halo model includes this absorption directly, whereas \citet{DallAglio2009} used the mode of the proximity-effect strength distribution to mitigate bias. Our central estimate lies above both the nearby \citet{BeckerBolton2013} measurements (whose error bars overlap our full interval) and the \citet{FaucherGiguere2020} model curve. We defer measurements at other redshifts to a subsequent paper.

The solid error bars in Figure~\ref{fig:evolution} show statistical uncertainties estimated by bootstrap resampling the DESI sightlines. These solid error bars are most directly comparable to those of \citet{DallAglio2009}, who also estimated their uncertainties by bootstrap resampling the sightlines. Our larger quasar sample yields smaller statistical errors than theirs in nearby redshift bins. The dashed error bars additionally include the modeled systematic uncertainties described in Section~\ref{sec:systematics}, showing how systematic uncertainty limits the precision of the UVBG inference. We discuss this comparison further in Section~\ref{sec:discussion}.

\FloatBarrier

\section{Uncertainty analysis and propagation}
\label{sec:systematics}

We report statistical errors estimated by bootstrap resampling the DESI sightlines separately from the systematic uncertainty. The latter includes uncertainties in the modeled transmission profile, continuum correction, mock calibration, external quasar bias, ionizing spectral slope, mean forest opacity, and quasar redshifts. We propagate the model-error floor, continuum-correction, mock-calibration, and quasar-bias uncertainties through the analysis, and estimate the mean-opacity, spectral-slope, and redshift contributions by varying these inputs and measuring the resulting changes in the inferred UVBG (Appendices~\ref{app:meanflux_sensitivity}, \ref{app:sed_sensitivity}, and \ref{app:redshift_offset}, respectively). Table~\ref{tab:uncertainty_budget} summarizes these contributions, and Section~\ref{sec:expanded_budget} describes how we combine them.

\begin{table*}[t]
\centering
\caption{Sources of uncertainty in the UVBG measurement and how we account for them. Resampling the DESI data alone gives the statistical uncertainty. The remaining contributions enter the systematic uncertainty, combined as described in Section~\ref{sec:expanded_budget}. The appendix tests listed below also check how the results depend on choices in the analysis.}
\label{tab:uncertainty_budget}
\small
\renewcommand{\arraystretch}{1.2}
\begin{tabular*}{\textwidth}{@{\extracolsep{\fill}}p{0.26\textwidth}p{0.69\textwidth}@{}}
\toprule
Source of uncertainty & Treatment\\
\midrule
\multicolumn{2}{@{}l}{\textbf{Statistical uncertainty}}\\
\addlinespace[2pt]
DESI sightline sampling
& Bootstrap resample the DESI quasars and refit their stacked transmission. Hold each spectrum's mean-flux regulation correction, the mock-derived wavelength-dependent continuum correction, and the UVBG calibration fixed, while recalculating the final external-forest normalization for each resampled stack (Sections~\ref{sec:method} and \ref{sec:expanded_budget}).\\
\midrule
\multicolumn{2}{@{}l}{\textbf{Systematic uncertainty}}\\
\addlinespace[2pt]
\raggedright Uncertainty in the modeled transmission
& Add independent Gaussian offsets with standard deviation $\sigma_{\rm mod}=0.015$ to each DESI radial bin and refit the perturbed profiles during uncertainty propagation. We also include this floor in the covariance matrices used to fit both the DESI and mock transmission profiles (Sections~\ref{sec:model_error} and \ref{sec:correction_uncertainties}). Appendices~\ref{sec:floor_sensitivity} and \ref{sec:floor_diagnostic} test other floor values and explain the floor estimates, respectively.\\
\raggedright Mock-derived continuum correction
& Bootstrap resample DESI and mock quasars, recompute the mock-derived wavelength-dependent continuum correction and final stack normalization of Sections~\ref{sec:mock_continuum_correction}--\ref{sec:final_normalization}, and propagate their uncertainty as described in Section~\ref{sec:correction_uncertainties}. Appendices~\ref{app:template_groups} and \ref{app:mock_noise} test different groups of HST continuum templates and an independent pixel-noise realization, respectively.\\
\astrid{} UVBG calibration
& Bootstrap resample the mock sightlines and recompute the UVBG calibration of Section~\ref{sec:astrid_response}, using the same resampled sightlines at every input rate (Section~\ref{sec:correction_uncertainties}). Appendix~\ref{app:template_groups} tests this calibration with different groups of HST continuum templates.\\
External quasar bias
& Vary the four DESI clustering measurements of \citet{Charles2026} adopted in Section~\ref{sec:desi_bias_assignment} within their quoted errors and propagate the resulting uncertainty in $b_q$ into the inferred UVBG (Section~\ref{sec:external_quasar_bias}).\\
Ionizing spectral slope
& Vary the quasar ionizing spectral slope $\alpha$ defined in Section~\ref{sec:quasar_radiation}. Scale the resulting UVBG changes using the constraint from \citet{Telfer2002}, with \citet{Lusso2015} as an alternative (Section~\ref{sec:sensitivity_slope} and Appendix~\ref{app:sed_sensitivity}).\\
Mean forest opacity
& Vary the amplitude and redshift exponent of the mean forest opacity relation in Section~\ref{sec:model_extensions}. Scale the resulting UVBG changes using the uncertainties from \citet{Turner2024} (Section~\ref{sec:sensitivity_opacity} and Appendix~\ref{app:meanflux_sensitivity}).\\
Quasar-redshift errors
& Test a common shift in quasar redshifts and scale its UVBG effect using the mean-offset constraint from \citet{Bault2025}. For random redshift errors, draw random offsets using each quasar's DESI redshift uncertainty and measure their effect on the calibrated UVBG (Section~\ref{sec:sensitivity_redshifts} and Appendix~\ref{app:redshift_offset}).\\
\bottomrule
\end{tabular*}
\end{table*}

\subsection{Model error floor}
\label{sec:model_error}

Bootstrap resampling whole DESI quasar sightlines gives the covariance of the stacked transmission profile, including correlations between radial bins (Section~\ref{sec:method}). Evaluating the best-fitting model with this covariance alone gives a reduced $\chi^2=\ZThreeChiNoFloor$. Sightline sampling therefore does not account for the full difference between the measured and modeled profiles.

We add a model-error floor of $\sigma_{\rm mod}=0.015$ in quadrature to each bin's sampling uncertainty,
\begin{equation}
\mathbf C=\mathbf C_{\rm boot}+\sigma_{\rm mod}^{2}\mathbf I,
\label{eq:model_floor_covariance}
\end{equation}
where $\mathbf C_{\rm boot}$ is the bootstrap covariance after the 30 percent diagonal shrinkage described in Section~\ref{sec:method}. The floor corresponds to 1.5 percentage points of transmitted flux and is treated as uncorrelated between bins.  The floor accounts for differences between the predicted and measured transmission that are larger than expected from the DESI statistical uncertainty.

 We estimate the DESI floor from the differences between the measured and best-fitting transmission profiles, accounting for the statistical uncertainty and the fitted UVBG parameter (Appendix~\ref{sec:floor_diagnostic}). For the clustering-bias fit, this gives $\sigma_{\rm mod}=\ZThreeDESIFloorHat^{+0.0045}_{-0.0030}$, consistent with our adopted value of 0.015. The same calculation supports this choice for the profile-fit bias. Varying the DESI floor to 0.010 and 0.020 changes the central UVBG estimate by only a few percent, while changing its uncertainty more substantially (Appendix~\ref{sec:floor_sensitivity}).

 We adopt the same floor for the \astrid{} mocks, consistent with estimates near the measured UVBG (Appendix~\ref{sec:floor_amplitude}). Using 0.010 or 0.020 instead changes the primary UVBG estimate by at most 1.2 percent (Appendix~\ref{sec:floor_sensitivity}).

\subsection{Propagating uncertainty in the continuum correction and mock calibration}
\label{sec:correction_uncertainties}

The continuum correction and UVBG calibration are estimated from the same finite mock sample, so their uncertainties are linked. We measure their combined effect on the inferred UVBG by resampling the DESI and mock sightlines and repeating the analysis. This follows changes in the measured transmission, continuum correction, and calibration through to the final photoionization rate.

We create 512 bootstrap realizations by drawing whole sightlines with replacement from each sample. For each realization, we recalculate the common wavelength-dependent continuum correction from the resampled mocks and apply it to both the DESI and mock spectra. We rebuild the measured and theoretical transmission stacks and recalculate the final scalar normalization that matches the stacked transmission outside the proximity region to the adopted mean forest transmission. We do this separately for the resampled DESI and mock samples, then refit the photoionization rates. Comparing the recovered mock rates with their known inputs gives a new calibration curve, which we use to calibrate that realization's DESI fit. We use the same resampled mock sightlines at every injected rate so that changes along the calibration curve reflect a common sample.

We also estimate the DESI statistical uncertainty from these 512 resampled DESI samples, keeping the continuum correction and mock calibration fixed while recalculating the final external-forest normalization and refitting each stack.

We also propagate the DESI model-error floor from Section~\ref{sec:model_error}. For each bootstrap realization, we create 32 versions of the DESI stack by adding independent random offsets to the radial bins, with standard deviation $\sigma_{\rm mod}=0.015$. These offsets represent the additional transmission uncertainty allowed by the floor. We refit each version using the covariance in Section~\ref{sec:method} and apply the calibration curve from the same bootstrap realization. The resulting spread in calibrated rates therefore includes both sampling uncertainty and the adopted model-error allowance.

The continuum fits and external-forest polynomial corrections for individual spectra remain fixed during resampling. The normalization recalculated above is the final scalar correction to each stack. The full-sample pixel-clipping masks, mean-opacity relation, floor amplitudes, and mock bias also remain fixed. This calculation measures sampling uncertainty within the available sightlines and HST continuum templates.

After including the quasar-bias uncertainty described in Section~\ref{sec:external_quasar_bias}, we use the 16th and 84th percentiles of the resulting rate distribution to define an intermediate uncertainty interval. Section~\ref{sec:expanded_budget} describes how we use this interval to calculate the separately reported statistical and systematic uncertainties.

\subsection{Uncertainty in the DESI quasar bias}
\label{sec:external_quasar_bias}

Uncertainty in the DESI quasar bias changes the predicted environmental absorption and therefore the photoionization rate inferred from the transmission profile. To include this contribution in the clustering-based UVBG measurement, we vary the four bias measurements of \citet{Charles2026} used to construct our redshift-dependent bias relation (Section~\ref{sec:desi_bias_assignment}).

We use a local linear response to avoid rebuilding and refitting the theoretical stack for every bias realization. We first measure the response of the fitted UVBG to each clustering measurement by shifting its bias up and down by one quoted standard deviation while holding the other three at their central values. For each shift, we rebuild the bias relation and theoretical transmission profiles and refit the DESI stack. Half the difference between the two fitted logarithmic UVBG rates estimates the response to a one-standard-deviation change. We then propagate the quoted clustering uncertainties by drawing independent standard Gaussian deviations for all four measurements in each uncertainty realization. Multiplying each deviation by its measured response and summing the four terms gives a perturbation to the fitted logarithmic UVBG rate, assuming a locally linear response. These draws represent the quoted statistical errors of the clustering measurements. We apply the perturbation before converting the rate through that realization's mock calibration. This propagates the quasar-bias uncertainty jointly with the contributions described in Section~\ref{sec:correction_uncertainties}, giving the intermediate UVBG uncertainty interval used in Section~\ref{sec:expanded_budget}.

\subsection{Ionizing-spectrum, mean-opacity, and redshift uncertainties}
\label{sec:sensitivity_contributions}

We estimate the ionizing-spectrum, mean-opacity, and redshift contributions from separate sensitivity tests to reduce the computational cost of repeating these variations within every bootstrap realization. These tests follow the effect of each variation through the relevant analysis steps, including recalculating the mock calibration for the mean-opacity and random-redshift tests. We convert the resulting changes into logarithmic uncertainty widths and add their variances to the uncertainty jointly propagated in Sections~\ref{sec:correction_uncertainties} and \ref{sec:external_quasar_bias} (Section~\ref{sec:expanded_budget}). This approximation neglects covariance among the additional contributions and with the jointly propagated terms.

\subsubsection{Mean forest opacity}
\label{sec:sensitivity_opacity}

The mean forest opacity sets the target transmission for continuum regulation (Section~\ref{sec:meanflux}). Using the amplitude and redshift-dependence tests in Appendix~\ref{app:meanflux_sensitivity}, we propagate the parameter uncertainties reported by \citet{Turner2024}, allowing for correlation between them. This gives a mean-opacity uncertainty of 0.008 dex in $\log_{10}\Gamma_{12}$.

\subsubsection{Ionizing spectral slope}
\label{sec:sensitivity_slope}

The ionizing spectral slope enters the luminosity-to-radiation conversion in Section~\ref{sec:quasar_radiation} and the equality radius in Equation~\ref{eq:req}. Using the tests in Appendix~\ref{app:sed_sensitivity}, we propagate the radio-quiet constraint $\alpha=1.57\pm0.17$ from \citet{Telfer2002}, including its offset from our adopted $\alpha=1.5$. This gives a spectral-slope uncertainty of 0.048 dex in $\log_{10}\Gamma_{12}$.

The broader constraint $\alpha=1.70\pm0.61$ from \citet{Lusso2015} instead gives 0.169 dex, assuming the measured response extends beyond the tested slope range. These alternative estimates are not combined.

\subsubsection{Quasar redshift uncertainties}
\label{sec:sensitivity_redshifts}

Quasar-redshift errors affect continuum reconstruction (Section~\ref{sec:continuum_desi}) and the inferred quasar--absorber separation used in the profile fits (Section~\ref{sec:stack_construction}). Using the tests in Appendix~\ref{app:redshift_offset}, we propagate the common redshift-offset constraint from \citet{Bault2025}, giving an uncertainty of 0.145 dex in $\log_{10}\Gamma_{12}$. Random redshift errors drawn from the DESI catalog uncertainties give a smaller contribution of 0.0088 dex. Table~\ref{tab:systematic_contributions} summarizes the contributions combined in Section~\ref{sec:expanded_budget}.

\begin{table}[t]
\centering
\caption{Additional contributions to the systematic uncertainty for the primary measurement using the clustering quasar bias, derived from the tests in Section~\ref{sec:sensitivity_contributions}. Values are widths in $\log_{10}\Gamma_{12}$. The spectral-slope constraints from \citet{Telfer2002} and \citet{Lusso2015} are alternatives and are not combined.}
\label{tab:systematic_contributions}
\small
\begin{tabular}{@{}lr@{}}
\toprule
Source & Uncertainty (dex)\\
\midrule
Ionizing slope: Telfer (adopted) & 0.048\\
Ionizing slope: Lusso (comparison) & 0.169\\
Mean forest opacity & 0.008\\
Coherent redshift offset & 0.145\\
Random redshift errors & 0.0088\\
\bottomrule
\end{tabular}
\end{table}

\subsection{Combining statistical and systematic uncertainty}
\label{sec:expanded_budget}

The calculation in Sections~\ref{sec:correction_uncertainties} and \ref{sec:external_quasar_bias} combines DESI statistical uncertainty with the model-error, continuum-correction, mock-calibration, and quasar-bias contributions. We subtract the DESI statistical contribution and add the spectral-slope, mean-opacity, and redshift contributions calculated in Section~\ref{sec:sensitivity_contributions} to report statistical and systematic uncertainties separately. The dashed outer error bars in Figure~\ref{fig:evolution} show the total uncertainty, combining both.

We combine uncertainties in the logarithm of the photoionization rate. We subtract the squared DESI statistical uncertainty from the squared uncertainty calculated in Sections~\ref{sec:correction_uncertainties} and \ref{sec:external_quasar_bias}, then add the squared contributions from the spectral slope, mean opacity, and redshift errors:
\begin{equation}
s_{\rm sys}^{2}=s_{\rm base}^{2}-s_{\rm stat}^{2}+\sum_k s_k^2.
\label{eq:expanded_budget}
\end{equation}
Here $s_{\rm base}$ is the uncertainty from Sections~\ref{sec:correction_uncertainties} and \ref{sec:external_quasar_bias}, $s_{\rm stat}$ is the DESI bootstrap statistical uncertainty, and $s_k$ are the contributions calculated in Section~\ref{sec:sensitivity_contributions}, all expressed in $\ln\Gamma_{12}$. We perform this calculation separately for the lower and upper errors and convert the results back to photoionization-rate uncertainties. For a central rate $\Gamma_0$ and percentile bounds $\Gamma_{16}$ and $\Gamma_{84}$, the logarithmic widths are $\ln(\Gamma_0/\Gamma_{16})$ and $\ln(\Gamma_{84}/\Gamma_0)$. This approximate combination treats the added contributions as independent. We convert the widths in Table~\ref{tab:systematic_contributions} from dex to natural logarithms before using Equation~\ref{eq:expanded_budget}.

For the profile-fit quasar-bias result, we use its own spectral-slope and mean-opacity responses and the same fractional redshift contributions. We also perform sensitivity tests using different groups of HST continuum templates, an independent mock pixel-noise realization, alternative mappings of $r_{\rm eq}$, and different model-error floor amplitudes (Appendix~\ref{app:sensitivity_tests}). These tests check how the results depend on our analysis choices. Their shifts are not added as independent uncertainty contributions. Variation among the available mock spectra and the adopted model-error floor already enter the propagation in Section~\ref{sec:correction_uncertainties}.

Table~\ref{tab:allz} reports the statistical and systematic uncertainties separately. Combining their squared logarithmic widths and converting back to the rate gives a total uncertainty of $^{+\ZThreeClusteringExpandedTotalPlus}_{-\ZThreeClusteringExpandedTotalMinus}$ in $\Gamma_{12}$. Using the wider spectral-slope constraint of \citet{Lusso2015} instead of \citet{Telfer2002} gives a systematic uncertainty of $^{+1.142}_{-0.625}$, with all other contributions unchanged.

\section{Summary and discussion}
\label{sec:summary_discussion}

\subsection{Summary}
\label{sec:summary}

We have measured the hydrogen photoionization rate of the UVBG from the line-of-sight quasar proximity effect in DESI DR1. Our main methods and results are as follows.

\begin{enumerate}[label=(\alph*),leftmargin=*,itemsep=4pt]
\item We select \ZThreeNDESI{} quasars over $2.75\leq z<3.25$, with $M_{1450}\leq-25.5$ and red-side $S/N\geq3$. We reconstruct their continua using an HST-trained PCA model and combine their spectra into a stacked transmission profile.

\item We fit this profile with the halo model of \citet{Croft2025}, which describes both the transmission enhancement from quasar radiation and the additional absorption in the surrounding overdense gas. Independent clustering measurements constrain the quasar bias, allowing us to infer the UVBG from the transmission profile.

\item We calibrate the measurement using \ZThreeNASTRID{} accepted mock spectra constructed from \astrid{} absorption profiles, HST continuum templates, and DESI-like noise. We apply the same continuum corrections and profile-fitting procedure to the mocks and DESI. Comparing recovered and injected photoionization rates gives the calibration used to correct the DESI estimate.

\item We obtain
\[
\GammaHI=\left(\ZThreeClusteringGamma^{+\ZThreeClusteringSamplingPlus}_{-\ZThreeClusteringSamplingMinus}\right)\times10^{-12}\,\mathrm{s}^{-1}
\]
 at $z\simeq3$. The quoted statistical errors are estimated by bootstrap resampling the DESI sightlines. The systematic uncertainty is $^{+\ZThreeClusteringSystematicPlus}_{-\ZThreeClusteringSystematicMinus}\times10^{-12}\,\mathrm{s}^{-1}$. Using the bias fitted to the mock transmission profile gives $\Gamma_{12}=\ZThreeEffectiveGamma$, with a central estimate only 3.3 percent higher.
\end{enumerate}

\subsection{Discussion}
\label{sec:discussion}

To our knowledge, this is the first measurement of the UVBG from the line-of-sight quasar proximity effect in DESI. With \ZThreeNDESI{} quasars over $2.75\leq z<3.25$, our sample size substantially exceeds those of previous proximity-effect measurements near $z=3$, enabling a precise measurement within a narrow redshift interval. The halo model reproduces the overall transmission profiles of the matched \astrid{} mocks and provides good fits to the DESI profile, with reduced $\chi^2$ values of \ZThreeClusteringChi{} and \ZThreeEffectiveChi{} for the two quasar-bias choices. The calibrated UVBG estimate is largely insensitive to most of the tested modeling and analysis assumptions, with moderate sensitivity to the ionizing spectral slope and coherent quasar-redshift offsets (Appendix~\ref{app:sensitivity_tests}). The DESI redshift offsets measured by \citet{Bault2025} near our sample redshift are small and consistent with zero. Nonetheless, we propagate the quoted uncertainty in the mean offset into the systematic uncertainty of our UVBG measurement. The mock calibration brings the UVBG estimates obtained with the clustering and profile-fit quasar biases into agreement to within 3.3 percent, as expected if the mocks and DESI data respond similarly to changes in the assumed quasar bias. This agreement supports the ability of the mocks to reproduce that response and calibrate the measurement.

Our analysis advances proximity-effect measurements by explicitly modeling absorption in the overdense environments of quasars, with the environmental contribution constrained by independent clustering measurements. Our central UVBG estimate is approximately 17 percent below the overall level $\Gamma_{12}\simeq1.66$ reported by \citet{DallAglio2009}, with overlapping uncertainties. This lower value is consistent with the expected effect of accounting for environmental absorption, which can otherwise bias proximity-effect estimates high \citep{LoebEisenstein1995,FaucherGiguere2008}. We also develop a continuum reconstruction method tailored to the proximity region. Previous analyses estimated continua using spline fits through the absorption region \citep{DallAglio2009,Jalan2021}, whereas our PCA prediction uses only red-side spectral information, with additional adjustments to the Ly$\alpha$ emission profile determined from its red wing. We further refine these predictions using the mean transmitted flux in the forest outside the proximity region and corrections derived from mock spectra. Neither the initial continuum prediction nor the subsequent corrections use the observed transmission in the proximity region to fit the continuum, preserving the signal we seek to measure.

 The large DESI sample provides substantially smaller statistical errors than the bootstrap uncertainties reported by \citet{DallAglio2009} in nearby redshift bins. Our statistical uncertainties are shown by the solid inner error bars in Figure~\ref{fig:evolution}. Our estimated systematic uncertainty is about eight times the statistical uncertainty in logarithmic width, so the UVBG measurement is limited primarily by continuum reconstruction, model calibration, the ionizing spectrum, and redshift uncertainties. With independent constraints on the UVBG, the measured proximity profiles could constrain the conversion from quasar luminosity to ionizing radiation and test descriptions of quasar environments. 

Reducing the systematic uncertainty will require better continuum predictions and a more accurate conversion from observed quasar luminosity to ionizing radiation. A continuum training sample spanning a wider range of quasar luminosities would improve predictions for the DESI population and extend the method to fainter quasars, whose weaker proximity effects require more accurate continuum reconstruction. Better measurements of quasar ionizing spectra would reduce uncertainty in the radiation reaching the absorbing gas. Extending the analysis beyond $z\simeq3$ will also require matched mock spectra that test continuum reconstruction and UVBG recovery as forest absorption and quasar environments evolve. These developments would enable measurements of the UVBG across a wider redshift range, which we plan to pursue in a subsequent paper.

Beyond measuring the mean UVBG and its evolution with redshift, a longer-term goal is to estimate the local UVBG from the proximity effect in individual quasar spectra. As explored by \citet{Croft2025}, these measurements could sample the radiation field at quasar positions throughout the DESI survey volume. With sufficient precision, combining them could reveal spatial fluctuations and measure the large-scale structure of the ionizing radiation field. The stacked measurement and mock validation presented here provide a foundation for developing this future application.

\section*{Data and code availability}

The DESI DR1 spectra used in this work are publicly available at
\url{https://data.desi.lbl.gov/public/dr1/}.
The analysis code, mock spectra, and derived data products underlying
the results presented here are available from the corresponding
author upon request.

\section*{Acknowledgments}

We thank Wynne Turner for helpful correspondence about LyCAN and
quasar continuum reconstruction, and Mikel Charles for useful
discussions about quasar clustering and bias.

This research used data obtained with the Dark Energy Spectroscopic Instrument (DESI). DESI construction and operations is managed by the Lawrence Berkeley National Laboratory. This material is based upon work supported by the U.S. Department of Energy, Office of Science, Office of High-Energy Physics, under Contract No. DE--AC02--05CH11231, and by the National Energy Research Scientific Computing Center, a DOE Office of Science User Facility under the same contract.

Additional support for DESI was provided by the U.S. National Science Foundation (NSF), Division of Astronomical Sciences under Contract No. AST-0950945 to the NSF's National Optical-Infrared Astronomy Research Laboratory; the Science and Technology Facilities Council of the United Kingdom; the Gordon and Betty Moore Foundation; the Heising-Simons Foundation; the French Alternative Energies and Atomic Energy Commission (CEA); the National Council of Humanities, Science and Technology of Mexico (CONAHCYT); the Ministry of Science and Innovation of Spain (MICINN), and by the DESI Member Institutions: \url{https://www.desi.lbl.gov/collaborating-institutions}.

The DESI collaboration is honored to be permitted to conduct scientific research on I'oligam Du'ag (Kitt Peak), a mountain with particular significance to the Tohono O'odham Nation. Any opinions, findings, and conclusions or recommendations expressed in this material are those of the author(s) and do not necessarily reflect the views of the U.S. National Science Foundation, the U.S. Department of Energy, or any of the listed funding agencies.

The DESI Legacy Imaging Surveys consist of three individual and complementary projects: the Dark Energy Camera Legacy Survey (DECaLS), the Beijing-Arizona Sky Survey (BASS), and the Mayall z-band Legacy Survey (MzLS). DECaLS, BASS and MzLS together include data obtained, respectively, at the Blanco telescope, Cerro Tololo Inter-American Observatory, NSF's NOIRLab; the Bok telescope, Steward Observatory, University of Arizona; and the Mayall telescope, Kitt Peak National Observatory, NOIRLab.

NOIRLab is operated by the Association of Universities for Research in Astronomy (AURA) under a cooperative agreement with the National Science Foundation. Pipeline processing and analyses of the data were supported by NOIRLab and the Lawrence Berkeley National Laboratory (LBNL).

Legacy Surveys also uses data products from the Near-Earth Object Wide-field Infrared Survey Explorer (NEOWISE), a project of the Jet Propulsion Laboratory/California Institute of Technology, funded by the National Aeronautics and Space Administration. Legacy Surveys was supported by: the Director, Office of Science, Office of High Energy Physics of the U.S. Department of Energy; the National Energy Research Scientific Computing Center, a DOE Office of Science User Facility; the U.S. National Science Foundation, Division of Astronomical Sciences; the National Astronomical Observatories of China, the Chinese Academy of Sciences and the Chinese National Natural Science Foundation. LBNL is managed by the Regents of the University of California under contract to the U.S. Department of Energy. The complete acknowledgments can be found at \url{https://www.legacysurvey.org/acknowledgment/}.

This research uses services or data provided by the Astro Data Lab at NSF's NOIRLab. NOIRLab is operated by the Association of Universities for Research in Astronomy (AURA), Inc. under a cooperative agreement with the National Science Foundation.

This research has made use of the HSLA database, developed and maintained at STScI, Baltimore, USA.

The \astrid{} simulation was carried out using the Frontera
supercomputer at the Texas Advanced Computing Center.

\noindent\begin{minipage}{\columnwidth}
We used GPT-based tools from OpenAI and Claude
(Anthropic) to assist with manuscript editing and the development of
analysis and plotting code. The analysis was checked through the sensitivity and robustness tests described in the main text and appendices. All text, code, and calculations produced with the aid of these tools were carefully checked and validated by the authors, who take full responsibility for the analysis and results.
\end{minipage}\par

\FloatBarrier

\appendix

\renewcommand{\theHtable}{appendix.\thesection.\arabic{table}}
\renewcommand{\theHequation}{appendix.\thesection.\arabic{equation}}

\section{Sensitivity tests}
\label{app:sensitivity_tests}

We test how the inferred UVBG changes when we vary the inputs and assumptions used in the analysis. These sensitivity tests vary the continuum templates, pixel-noise realization, model-error floor, mean forest opacity, ionizing spectral slope, equality-radius mapping, and quasar redshifts. We compare the resulting UVBG estimates with the adopted measurement or, for mock recovery tests, with the known input rates.

\subsection{Transfer between continuum-template groups}
\label{app:template_groups}
We divide the continuum shapes derived from HST spectra and used to construct the mocks (Section~\ref{sec:mock_spectra}) into two disjoint groups. We construct the calibration using mocks assigned continuum shapes from one group, apply it to mocks assigned continuum shapes from the other, and then exchange the groups. The recovery errors are 0.0246--0.0285 dex for the clustering bias and 0.0251--0.0267 dex for the profile-fit bias. This tests transfer between the available continuum shapes within the same simulation. Table~\ref{tab:calibration_checks} gives the recovery errors.

\begin{table}[!htbp]
\centering
\caption{Mock-calibration recovery errors for transfer between continuum-template groups, expressed as root-mean-square differences in $\log_{10}\Gamma_{12}$ (dex). The arrows identify which template group supplies the calibration and which is used for testing. See Appendix~\ref{app:template_groups}.}
\label{tab:calibration_checks}
\small
\begin{tabular}{@{}lrr@{}}
\toprule
Test & Clustering & Profile-fit\\
\midrule
Templates: group 1 $\to$ 2 & 0.0246 & 0.0251\\
Templates: group 2 $\to$ 1 & 0.0285 & 0.0267\\
\bottomrule
\end{tabular}
\par\vspace{10pt}
\end{table}

\subsection{Dependence on the pixel-noise realization}
\label{app:mock_noise}
We test mock-calibration recovery with a second independent realization of the Gaussian noise added to the flux in each spectral pixel of the mock spectra, using the same pixel variances as in the main calibration (Section~\ref{sec:mock_spectra}). Table~\ref{tab:mock_noise} gives the root-mean-square recovery errors in $\log_{10}\Gamma_{12}$. For comparison, the primary measurement's combined statistical and systematic uncertainty has a logarithmic half-width of 0.208 dex.

\begin{table}[!htbp]
\centering
\caption{Mock-calibration recovery errors with an independent pixel-noise realization, expressed as root-mean-square differences in $\log_{10}\Gamma_{12}$ (dex). See Appendix~\ref{app:mock_noise}.}
\label{tab:mock_noise}
\begin{tabular}{@{}lr@{}}
\toprule
Bias choice & Recovery error (dex)\\
\midrule
Clustering & 0.0039\\
Profile-fit & 0.0048\\
\bottomrule
\end{tabular}
\end{table}

\subsection{Sensitivity to the model-error floor}
\label{sec:floor_sensitivity}

 The model-error floor is an additional uncertainty in transmitted flux added in quadrature to the sampling uncertainty in each radial bin to account for discrepancies between the halo model and the DESI or mock transmission profiles (Section~\ref{sec:model_error}). We test floor amplitudes of 0.010 and 0.020 in absolute transmission units for both DESI and the mocks, bracketing the adopted value of 0.015. We vary one floor at a time while holding the other at 0.015, then repeat the calibration and uncertainty propagation. The spectra, continuum fits, and quasar-bias choices remain fixed. Table~\ref{tab:floor_sensitivity} gives the results for the clustering quasar bias.

 Changing the mock floor changes either the upper or lower error by at most 1.7 percent. The mock floor controls the weights used to fit the calibration profiles, whereas the DESI floor also sets the size of the transmission perturbations used in uncertainty propagation (Section~\ref{sec:correction_uncertainties}).

\begin{table}[!htbp]
\centering
\caption{Sensitivity of the calibrated UVBG measurement to the DESI and \astrid{} model-error floors, using the clustering quasar bias. The quoted errors come from the 16th and 84th percentiles of the rate distribution calculated in Sections~\ref{sec:correction_uncertainties} and \ref{sec:external_quasar_bias}. They include DESI statistical uncertainty and the propagated model-error, continuum-correction, mock-calibration, and quasar-bias contributions. Our primary measurement adopts 0.015 for both floors. See Appendix~\ref{sec:floor_sensitivity}.}
\label{tab:floor_sensitivity}
\begin{tabular}{@{}ccc@{}}
\toprule
DESI floor & \astrid{} floor & Calibrated $\Gamma_{12}$\\
\midrule
0.015 & 0.015 & $\ZThreeClusteringGamma^{+\ZThreeClusteringPlus}_{-\ZThreeClusteringMinus}$\\
0.015 & 0.010 & ${1.391}^{+0.537}_{-0.383}$\\
0.015 & 0.020 & ${1.369}^{+0.525}_{-0.380}$\\
0.010 & 0.015 & ${1.409}^{+0.423}_{-0.335}$\\
0.020 & 0.015 & ${1.363}^{+0.682}_{-0.437}$\\
\bottomrule
\end{tabular}
\end{table}

\subsection{Sensitivity to the mean forest opacity}
\label{app:meanflux_sensitivity}

We repeat the continuum regulation and UVBG calibration after varying the mean-opacity relation defined in Section~\ref{sec:model_extensions} and used to regulate the continua of both DESI and the mocks (Section~\ref{sec:meanflux}). The amplitude tests multiply $\tau_0$ by 0.95 or 1.05 at fixed exponent $\beta=3.62$. The exponent tests use $\beta=3.52$ or 3.72 and adjust $\tau_0$ to keep $\bar\tau_{\rm eff}(3)$ unchanged. The sample remains fixed. Table~\ref{tab:opacity_tests} lists the resulting clustering-based UVBG estimates. Their changes are at most 1.2 percent.

To calculate the mean-opacity uncertainty reported in Section~\ref{sec:sensitivity_opacity}, we write $\bar\tau_{\rm eff}(z)=\tau_0(1+z)^\beta$ and scale the measured responses using $\tau_0=(2.46\pm0.14)\times10^{-3}$ and $\beta=3.62\pm0.04$ from \citet{Turner2024}. Because the exponent tests preserve the opacity at $z=3$, we use
\begin{equation}
\delta\ln\bar\tau_{\rm eff}(3)=\delta\ln\tau_0+\ln(4)\delta\beta
\end{equation}
to express the response in terms of the published parameters $\tau_0$ and $\beta$. We add the magnitudes of their uncertainty contributions, allowing for correlation between their errors. This gives a mean-opacity uncertainty of 0.008 dex in $\log_{10}\Gamma_{12}$.

\begin{table}[!htbp]
\centering
\caption{Sensitivity to mean forest opacity, using the clustering quasar bias. Changes are relative to the fiducial result. The exponent variations preserve the opacity at $z=3$. See Appendix~\ref{app:meanflux_sensitivity}.}
\label{tab:opacity_tests}
\small
\begin{tabular}{@{}lrr@{}}
\toprule
Opacity choice & $\Gamma_{12}^{\rm cal}$ & Change (\%)\\
\midrule
Fiducial & 1.376 & 0.00\\
$0.95\tau_0$ & 1.388 & $+0.88$\\
$1.05\tau_0$ & 1.369 & $-0.49$\\
$\beta=3.52$ & 1.387 & $+0.80$\\
$\beta=3.72$ & 1.360 & $-1.19$\\
\bottomrule
\end{tabular}
\end{table}

\subsection{Sensitivity to the ionizing spectral slope}
\label{app:sed_sensitivity}

We vary the ionizing spectral slope in the DESI luminosity-to-radiation conversion described in Section~\ref{sec:quasar_radiation} and refit the DESI profile. The radiation injected into the mocks is known, so its value and the corresponding calibration curve remain fixed in this test. Table~\ref{tab:slope_tests} gives the clustering-based estimates.

To obtain the spectral-slope uncertainty reported in Section~\ref{sec:sensitivity_slope}, we estimate the response $D_\alpha=\partial\ln\Gamma_{12}/\partial\alpha$ from the tests at $\alpha=1.3$ and $1.7$. For a published slope constraint $\alpha_{\rm obs}\pm\sigma_\alpha$ and our adopted slope $\alpha_0=1.5$, we calculate the uncertainty width
\begin{equation}
s_\alpha=|D_\alpha|\sqrt{\sigma_\alpha^2+(\alpha_{\rm obs}-\alpha_0)^2}.
\end{equation}
Here $s_\alpha$ is expressed in $\ln\Gamma_{12}$. This combines the quoted slope uncertainty and its offset from our adopted value in quadrature. Dividing by $\ln 10$ gives the width in dex. The radio-quiet constraint $\alpha_{\rm obs}=1.57$ and $\sigma_\alpha=0.17$ from \citet{Telfer2002} gives 0.048 dex. Using $\alpha_{\rm obs}=1.70$ and $\sigma_\alpha=0.61$ from \citet{Lusso2015} gives 0.169 dex, assuming the response remains linear beyond the tested interval. We adopt the Telfer estimate and retain the Lusso estimate as an alternative.

\begin{table}[!htbp]
\centering
\caption{Sensitivity to the ionizing spectral slope, using the clustering quasar bias. Changes are relative to the adopted $\alpha=1.5$. See Appendix~\ref{app:sed_sensitivity}.}
\label{tab:slope_tests}
\begin{tabular}{@{}crr@{}}
\toprule
$\alpha$ & $\Gamma_{12}^{\rm cal}$ & Change (\%)\\
\midrule
1.3 & 1.207 & $-12.30$\\
1.5 & 1.376 & 0.00\\
1.7 & 1.538 & $+11.78$\\
\bottomrule
\end{tabular}
\end{table}

\subsection{Mapping the equality radius to the halo-model grid}
\label{app:radius_mapping}

We determine the relation between the known equality radius $r_{\rm eq}$ in the simulations and the fitted halo-model radius $r_{\rm grid}$. Section~\ref{sec:method} describes how this mapping enters the profile fits.

Our analysis adopts the mapping obtained by fitting the halo-model grid to digitized halo-model curves shown alongside the \astrid{} profiles in Figure~7 of \citet{Croft2025}. These curves correspond to known equality radii of 0, 1.25, 2.5, 5, 10, and $20\,h^{-1}\,\mathrm{pMpc}$. We fitted the relation between the known radii and the recovered model-grid radii with a power law plus a constant offset. Expressing both radii in units of $h^{-1}\,\mathrm{pMpc}$ gives
\begin{equation}
 r_{\rm grid}=1.63867\,r_{\rm eq}^{0.824167}+0.144796.
\end{equation}

We also test a mapping fitted directly to the simulated transmission. This directly fitted relation is an alternative to the adopted mapping. We generated profiles from the original optical-depth, velocity, and temperature fields of 1,000 \astrid{} sightlines, assigning the same known equality radius to every quasar in each realization. We applied the quasar radiation before peculiar velocities and thermal broadening, following Section~\ref{sec:astrid_hybrid_mocks}. At each of the six radii listed above, we averaged the transmission profiles and fitted the model-grid radius using equal weights for the 50 native velocity bins spanning $100$--$9900\kms$, with the quasar bias set to the clustering-reference value $b_q=5.593$. These fits use the native profiles before continuum reconstruction and mean-opacity evolution. Fitting the recovered radii with the same functional form gives
\begin{equation}
 r_{\rm grid}=0.993092\,r_{\rm eq}^{0.959026}+1.000044.
\end{equation}

Although the fitted coefficients differ, the two relations give similar model radii over the range relevant to our DESI sample. At the reference rate $\Gamma_{12}=1.376$, the quasars have equality radii spanning $1.47$--$6.09\,h^{-1}\,\mathrm{pMpc}$. Evaluating both relations at each quasar's radius gives a mean absolute fractional difference of 2.7 percent relative to the adopted mapping, with a maximum difference of 10.7 percent across this full radius range.

We repeat the UVBG analysis with the directly fitted relation and with no remapping, $r_{\rm grid}=r_{\rm eq}$. In each case, we refit DESI and mock profiles, rebuild the calibration, and propagate the uncertainty. Table~\ref{tab:radius_mapping_tests} gives the results. The clustering-based central rates change by at most 2.0 percent and their upper and lower errors by less than 4 percent. The fitted mock bias changes more substantially, consistent with the mapping absorbing some of the mismatch between the halo model and the simulated transmission profiles.

\begin{table*}[!t]
\centering
\caption{Equality-radius mapping tests. UVBG estimates use the fixed clustering bias. Errors include the contributions propagated in Sections~\ref{sec:correction_uncertainties} and \ref{sec:external_quasar_bias}. The second column gives the bias fitted separately to the mock profile at injected $\Gamma_{12}=1$. Percentage changes in the calibrated UVBG are relative to the adopted digitized-curve mapping and calculated from the displayed central rates. See Appendix~\ref{app:radius_mapping}.}
\label{tab:radius_mapping_tests}
\small
\begin{tabular}{@{}lccr@{}}
\toprule
Mapping & Fitted $b_q$ & Calibrated $\Gamma_{12}$ & Change (\%)\\
\midrule
Digitized (adopted) & $\ZThreeEffectiveBias$ & $\ZThreeClusteringGamma^{+\ZThreeClusteringPlus}_{-\ZThreeClusteringMinus}$ & 0.0\\
Direct simulation fit & 3.405 & $1.367^{+0.520}_{-0.377}$ & $-0.7$\\
No remapping & 0.789 & $1.404^{+0.547}_{-0.392}$ & $+2.0$\\
\bottomrule
\end{tabular}
\end{table*}

\subsection{Sensitivity to quasar-redshift errors}
\label{app:redshift_offset}

Section~\ref{sec:sample_selection} describes the adopted quasar redshifts, and Section~\ref{sec:continuum_desi} describes their use in shifting the spectra to the rest frame.

\subsubsection{Coherent redshift offsets}
\label{app:coherent_redshift}
We impose common shifts of $\delta v=\pm10\kms$ on the assumed mock redshifts, where $\delta v=c(z_{\rm assumed}-z_{\rm true})/(1+z_{\rm true})$. At injected $\Gamma_{12}=1.25$, we resample the unchanged mock spectra using the shifted redshifts, propagate the interpolation variances, and repeat the continuum reconstruction, emission-line adjustments, external-forest correction, stacking, and profile fit. The sample, sightline weights, common wavelength correction, and UVBG calibration remain fixed. These tests use the saved rest-frame spectral grid.

Table~\ref{tab:coherent_redshift} gives the recovered rates. The offsets change the reconstructed continuum near the innermost fitted bin by about 1 percent, producing a larger fractional change in the inferred UVBG. To calculate the common-offset uncertainty reported in Section~\ref{sec:sensitivity_redshifts}, we use the $z>2.8$ updated-template result of \citet{Bault2025}, approximately $\mu_v=-12\kms$ with uncertainty $\sigma_v=15\kms$ at $z=2.98$, close to our sample redshift. For comparison, the broader DESI DR1 Ly$\alpha$ BAO sample gives approximately $7\pm6\kms$ \citep{DESI2024LyaBAO}. We estimate the response $D_v=\partial\ln\Gamma_{12}/\partial\delta v$ from the two offset tests and calculate the logarithmic uncertainty width as
\begin{equation}
s_v=|D_v|\sqrt{\sigma_v^2+\mu_v^2}.
\end{equation}
This includes both the uncertainty in the common offset and its displacement from zero. Dividing $s_v$ by $\ln 10$ gives 0.145 dex. This calculation assumes that the fractional response measured in the mocks applies to DESI and remains linear over the offsets implied by the published constraint.

\begin{table}[!htbp]
\centering
\caption{Recovery of injected $\Gamma_{12}=1.25$ with coherent quasar-redshift offsets. Changes are relative to the zero-offset control. See Appendix~\ref{app:coherent_redshift}.}
\label{tab:coherent_redshift}
\small
\begin{tabular}{@{}crr@{}}
\toprule
Offset ($\mathrm{km\,s^{-1}}$) & Recovered $\Gamma_{12}$ & Change (\%)\\
\midrule
$-10$ & 1.490 & $+19.2$\\
0 & 1.250 & 0.0\\
$+10$ & 1.053 & $-15.8$\\
\bottomrule
\end{tabular}
\end{table}

\subsubsection{Random redshift errors}
\label{app:random_redshift}
We obtain formal redshift uncertainties, \texttt{ZERR}, from the DESI DR1 Ly$\alpha$-optimized catalog for 5,219 selected quasars and from the primary DR1 catalog for the remaining 1,026, checking that the catalog redshifts match those used in the analysis.\footnote{The Ly$\alpha$-optimized catalog is available at \url{https://data.desi.lbl.gov/doc/releases/dr1/vac/zlya/}; primary values were retrieved from the NOIRLab Astro Data Lab table \texttt{desi\_dr1.zpix}.} We convert these to $\sigma_{v,i}=c\,\mathrm{ZERR}_i/(1+z_i)$. The median is $12.3\kms$, with 16th and 84th percentiles of 7.0 and $20.7\kms$; the sightline-weighted root-mean-square uncertainty is $14.6\kms$. These formal fitting errors do not capture persistent offsets caused by a mismatch between a quasar spectrum and its redshift template.

For each mock, we draw an offset from $\mathcal N(0,\sigma_{v,i}^2)$ using its matched DESI quasar's uncertainty. We generate three realizations and subtract the sightline-weighted mean offset from each to isolate random scatter. Each quasar retains the same offset across all 17 injected rates. We refit the continua and external-forest corrections, recompute the common wavelength correction for DESI and the mocks, rebuild the calibration, and refit DESI. The sample, weights, bias, and fit covariance remain fixed.

Table~\ref{tab:redshift_scatter} lists the resulting DESI estimates, including comparison tests with a common scatter of 25 or $50\kms$. The root-mean-square change in $\ln\Gamma_{12}$ across the three catalog-error realizations, relative to the zero-error control, is 0.0203, or 0.0088 dex. We adopt this as the random-redshift uncertainty reported in Section~\ref{sec:sensitivity_redshifts}.

\begin{table}[!htbp]
\centering
\caption{Sensitivity of the calibrated DESI UVBG estimate to random quasar-redshift offsets. Rows labeled DESI catalog use Gaussian offsets with a standard deviation set by the redshift uncertainty of each mock's matched DESI quasar. The other tests use the listed standard deviation for every quasar. Each realization is a new set of random offsets. Percentage changes are relative to the run with no added redshift offsets. See Appendix~\ref{app:random_redshift}.}
\label{tab:redshift_scatter}
\begin{tabular}{@{}rrrr@{}}
\toprule
Error model & Realization & $\Gamma_{12}^{\rm cal}$ & Change (\%)\\
\midrule
None & --- & 1.3760 & 0.0\\
\midrule
DESI catalog & 1 & 1.3942 & $+1.32$\\
DESI catalog & 2 & 1.4202 & $+3.21$\\
DESI catalog & 3 & 1.3868 & $+0.79$\\
\midrule
$25\kms$ & 1 & 1.423 & $+3.4$\\
$25\kms$ & 2 & 1.450 & $+5.4$\\
$25\kms$ & 3 & 1.404 & $+2.0$\\
\midrule
$50\kms$ & 1 & 1.537 & $+11.7$\\
$50\kms$ & 2 & 1.524 & $+10.8$\\
$50\kms$ & 3 & 1.521 & $+10.5$\\
\bottomrule
\end{tabular}
\end{table}

As an interpolation check, we reconstruct the mocks on an observed-wavelength grid and repeat the correction and calibration for three noise realizations. Changes relative to each realization's own zero-error control span approximately $-5$ to $+5$ percent for $25\kms$ scatter and $+3$ to $+17$ percent for $50\kms$ scatter. The reconstructed grid changes the zero-error baseline too, so these are numerical checks on the sensitivity test.

\FloatBarrier

\section{Estimating the model-error floor from profile residuals}
\label{sec:floor_diagnostic}

We test whether the residual differences between the measured transmission and the best-fitting halo-model transmission profile support the model-error floor adopted in Section~\ref{sec:model_error}. Section~\ref{sec:floor_amplitude} gives the floor amplitudes inferred from the DESI and mock residuals. Section~\ref{sec:floor_correlations} tests correlations between radial bins.

\subsection{Estimating the floor amplitude}
\label{sec:floor_amplitude}

The fit to the seven radial bins adjusts one parameter, $\GammaHI$. To estimate the additional uncertainty required by the remaining mismatch, we remove the component of the residuals that can be absorbed by a small change in this parameter. Let $\mathbf X$ be the vector of derivatives of the model transmission with respect to $\log\GammaHI$, evaluated at the best fit. We construct a matrix $\mathbf Q$ whose six orthonormal columns are perpendicular to $\mathbf X$, so that $\mathbf Q^{\mathsf T}\mathbf X=0$. The residuals and covariance in these six remaining directions are
\begin{equation}
\begin{aligned}
\mathbf r&=\mathbf Q^{\mathsf T}(\mathbf y-\mathbf m_{\rm best}),\\
\mathbf C_\perp&=\mathbf Q^{\mathsf T}\mathbf C\mathbf Q,
\end{aligned}
\end{equation}
where $\mathbf y$ is the measured transmission vector and $\mathbf m_{\rm best}$ is the transmission predicted by the best-fitting halo model.

For each trial floor amplitude, we use $\mathbf C=\mathbf C_{\rm boot}+\sigma_{\rm mod}^2\mathbf I$ and evaluate the Gaussian likelihood of these residuals,
\begin{equation}
-2\ln\mathcal L_\perp
=\mathbf r^{\mathsf T}\mathbf C_\perp^{-1}\mathbf r
+\ln\det\mathbf C_\perp+\mathrm{constant}.
\label{eq:floor_restricted_likelihood}
\end{equation}
The first term measures the mismatch relative to the trial uncertainty. Increasing the floor reduces this term, while the determinant term penalizes unnecessarily large uncertainties. Minimizing their sum therefore estimates the floor needed to describe the residuals. The quoted errors correspond to a rise of one in $-2\ln\mathcal L_\perp$ from its minimum.

We use the bootstrap covariance after diagonal shrinkage and hold it and $\mathbf Q$ fixed while varying the floor. For this diagnostic likelihood, we do not apply the Hartlap correction used in the profile fits. The resulting likelihood intervals describe a local approximation around the best-fitting profile, with the estimated sampling covariance treated as known. For DESI, the resulting estimates are $\ZThreeDESIFloorHat^{+0.0045}_{-0.0030}$ and $\ZThreeEffectiveFloorHat^{+0.0055}_{-0.0030}$ for the clustering and profile-fit biases, respectively (Table~\ref{tab:floor_residuals}). Both are consistent with 0.015. Applying the same amplitude calculation to the corrected \astrid{} profiles gives the estimates in Table~\ref{tab:mock_floor_amplitudes}. The best-fitting mock floors are close to the adopted value of 0.015 at injected $\Gamma_{12}=0.6$ and 1, while higher injected rates favor smaller floors. We propagate the adopted DESI floor through the transmission perturbations in Section~\ref{sec:correction_uncertainties}; Appendix~\ref{sec:floor_sensitivity} tests changes to either floor.

\begin{table}[!htbp]
\centering
\caption{Model-error floors estimated from the DESI transmission residuals. Errors correspond to a rise of one in $-2\ln\mathcal L_\perp$, with the bootstrap covariance and projection held fixed. The adopted floor is 0.015. See Appendix~\ref{sec:floor_amplitude}.}
\label{tab:floor_residuals}
\begin{tabular}{@{}lc@{}}
\toprule
Quasar-bias choice & Best-fitting $\sigma_{\rm mod}$\\
\midrule
Clustering & $\ZThreeDESIFloorHat^{+0.0045}_{-0.0030}$\\
Profile-fit & $\ZThreeEffectiveFloorHat^{+0.0055}_{-0.0030}$\\
\bottomrule
\end{tabular}
\end{table}

\begin{table}[!htbp]
\centering
\caption{Uncorrelated model-error floor amplitudes estimated from the corrected \astrid{} mock residuals, using the clustering quasar bias. The adopted floor is 0.015 at every injected rate. See Appendix~\ref{sec:floor_amplitude}.}
\label{tab:mock_floor_amplitudes}
\begin{tabular}{@{}cc@{}}
\toprule
Injected $\Gamma_{12}$ & Best-fitting $\sigma_{\rm mod}$\\
\midrule
0.6 & 0.0160\\
1 & 0.0135\\
2 & 0.0100\\
4 & 0.0060\\
8 & 0.0025\\
\bottomrule
\end{tabular}
\end{table}

\subsection{Correlations between radial-bin residuals}
\label{sec:floor_correlations}

The adopted floor adds independent uncertainty to each radial bin. To test whether the additional mismatch is better described by correlated errors, we replace $\mathbf I$ with
\begin{equation}
K_{ij}(\ell)=\exp\!\left(-\frac{|v_i-v_j|}{\ell}\right),
\label{eq:floor_correlation_kernel}
\end{equation}
where $v_i$ is the velocity of radial bin $i$. We test correlation lengths of $\ell=500$, 1500, and $3000\kms$, refitting the floor amplitude for each choice using Equation~\ref{eq:floor_restricted_likelihood}. The uncorrelated model gives the highest residual likelihood for DESI with both quasar-bias choices. This supports its use among the tested alternatives, although the six independent residual combinations provide limited information about the correlation pattern.

For the corrected \astrid{} mock profiles, correlated errors improve the description of the residuals. Thus the uncorrelated floor is an approximation for the mocks, even though it is preferred among the tested forms for DESI. Appendix~\ref{sec:floor_sensitivity} varies the floor amplitude while retaining the uncorrelated form.

These estimates describe residual mismatch beyond the sightline-bootstrap covariance, after removing the direction associated with changing the fitted UVBG. They therefore assess the remaining profile shape rather than the accuracy of the recovered photoionization rate. Recovery of the injected UVBG is tested separately through the mock calibration in Section~\ref{sec:astrid_response}.

\FloatBarrier

\bibliographystyle{mnras}

\bibliography{ref}

\end{document}